\documentclass{elsart}

\usepackage[square,comma]{natbib}
\usepackage{graphicx}
\usepackage{pxfonts}
\usepackage{lineno}

\usepackage{amssymb}
\usepackage{footnote}
\usepackage{multirow}
\usepackage{varwidth}
\usepackage{array}
\usepackage{epstopdf}
\usepackage{url}

\journal{}

\begin{document}


\thispagestyle{empty}
\begin{Large}
\textbf{DEUTSCHES ELEKTRONEN-SYNCHROTRON}

\textbf{\large{in der HELMHOLTZ-GEMEINSCHAFT}\\}
\end{Large}

DESY 15-141

August 2015

\begin{eqnarray}
\nonumber
\end{eqnarray}
\begin{center}
\begin{Large}
\textbf{Scheme to increase the output average spectral flux of the European XFEL at $14.4$ keV}
\end{Large}
\begin{eqnarray}
\nonumber &&\cr \nonumber && \cr
\end{eqnarray}

\begin{large}
Gianluca Geloni,
\end{large}
\textsl{\\European XFEL GmbH, Hamburg}
\begin{large}

Vitali Kocharyan and Evgeni Saldin
\end{large}
\textsl{\\Deutsches Elektronen-Synchrotron DESY, Hamburg}

\begin{eqnarray}
\nonumber
\end{eqnarray}
\begin{eqnarray}
\nonumber
\end{eqnarray}
ISSN 0418-9833
\begin{eqnarray}
\nonumber
\end{eqnarray}
\begin{large}
\textbf{NOTKESTRASSE 85 - 22607 HAMBURG}
\end{large}
\end{center}
\clearpage
\newpage
\begin{frontmatter}



\title{Scheme to increase the output average spectral flux of the European XFEL at $14.4$ keV}


\author[XFEL]{Gianluca Geloni,}
\author[DESY]{Vitali Kocharyan,}
\author[DESY]{Evgeni Saldin}
\address[XFEL]{European XFEL GmbH, Hamburg, Germany}
\address[DESY]{Deutsches Elektronen-Synchrotron (DESY), Hamburg, Germany}

\begin{abstract}
Techniques like inelastic X-ray scattering (IXS) and nuclear resonance scattering (NRS) are currently limited by the photon flux available at X-ray sources. At $14.4$ keV, third generation synchrotron radiation sources produce a maximum of $10^{10}$ photons  per second in a meV bandwidth.  In this work we discuss about the possibility of increasing this flux a thousand-fold by exploiting high repetition rate self-seeded pulses at the European XFEL. Here we report on a feasibility study for an optimized configuration of the SASE2 beamline at the European XFEL which combines self-seeding and undulator tapering techniques in order to increase the average spectral flux at $14.4$ keV. In particular, we propose to perform monochromatization at $7.2$ keV with the help of self-seeding, and amplify the seed in the first part of output undulator. The amplification process can be stopped at a position well before  saturation, where the electron beam gets considerable bunching at the 2nd harmonic of the coherent radiation. A second part of the output undulator follows, tuned to the 2nd harmonic frequency, i.e. at $14.4$ keV and is used to obtain saturation at this energy. One can further prolong the exchange of energy between the photon and the electron beam by tapering the last part of the output undulator. We performed start-to-end simulations and demonstrate that self-seeding, combined with undulator tapering, allows one to achieve more than a hundred-fold increase in average spectral flux compared with the nominal SASE regime at saturation, resulting in a maximum flux of order $10^{13}$ photons per second in a meV bandwidth.
\end{abstract}

%
%
%
\end{frontmatter}



\section{\label{sec:intro} Introduction and method}

Inelastic X-ray scattering from electrons \cite{BDP87}-\cite{Baron15}  (IXS) and Nuclear Resonant Scattering \cite{RUBY}-\cite{GER2} (NRS) are important techniques for probing condensed matter by successfully exploiting the high brightness of synchrotron radiation sources. Inelastic scattering relies on the transfer of momentum and energy from the photon field to the sample. Such transfer is detected as a change of momentum and energy of the scattered photons, and allows for the study of a number of excitations with different characteristic lengths and time scales, related to the momentum and energy transfer. This technique calls for a very high average spectral density of the incident X-ray radiation.

In \cite{OURIXS} it was studied how sub-meV inelastic X-ray scattering experiments could benefit from high-repetition rate, seeded XFELs. In that case, the method exploited Hard X-ray Self-Seeding (HXRSS) and tapering at the European XFEL around the photon energy of $9$ keV in combination with a new concept for monochromatization and spectral analysis \cite{SSM13, SSS13,Shvydko15}, which is expected to lead to Ultra-High Resolution IXS (UHRIX) momentum-resolved experiments with 0.1-meV spectral and 0.02-nm$^{-1}$ momentum transfer resolution. In that work we showed that the European XFEL equipped with HXRSS can lead to a photon flux of order $10^{14}$ ph/s/meV  at the sample, about four orders of magnitude larger than what is presently achievable at synchrotrons.

A continuation of that kind of studies includes the investigation of the average spectral density achievable at high-repetition self-seeded FEL at higher X-ray energies. The most natural application of these kind of X-ray sources is for NRS experiments. Similarly as for IXS applications, also NRS analyzers are limited, in resolution, by the flux available at synchrotrons. For instance, at $14.4$ keV, the maximum spectral flux available at third generation synchrotron radiation sources is of order $10^{10}$ photons per second per meV bandwidth.

Here we propose a way to increase such spectral flux up to about three orders of magnitude at the SASE2 beamline of the European XFEL. This will enable NRS experiments with very high, sub-meV resolution, and can be achieved by exploiting a combination of three different factors and techniques: first, the high-repetition rate of the European XFEL \cite{XFEL}; second, the HXRSS setup \cite{SELF}-\cite{INAG} that will be installed at SASE2 \cite{XFELSEED}; third, Coherent Harmonic Generation (CHG) \cite{SCHM}-\cite{SPEZ}; and finally, fourth, post-saturation tapering \cite{TAP1}-\cite{LAST}.

X-Ray Free Electron Lasers (XFELs) are capable of producing X-ray pulses with unprecedented power spectral density. However, the average spectral flux strongly depends on the maximal repetition rate that can be achieved by the linac driver of each XFEL setup. In particular, the European XFEL \cite{XFEL} will be driven by a superconducting accelerator, which enables up to 27000 pulses per second, more than two orders of magnitude higher than what can be achieved with a normal-conducting linac. A straightforward analysis (see section \ref{fels}) for the case of the SASE2 beamline of the European XFEL at $14.4$ keV shows that one can obtain up to about $10^{11} \mathrm{ph}/\mathrm{s}/\mathrm{meV}$ for the SASE case at saturation. This number is already one order of magnitude better than what can be achieved at synchrotrons that can  provide about $10^{10}$ ph/s/meV around $14.4$ keV. However, the average spectral flux can be further increased of other two orders of magnitude by combining together HXRSS, CHG, and post-saturation tapering techniques.

\begin{figure}
\includegraphics[width=1.0\textwidth]{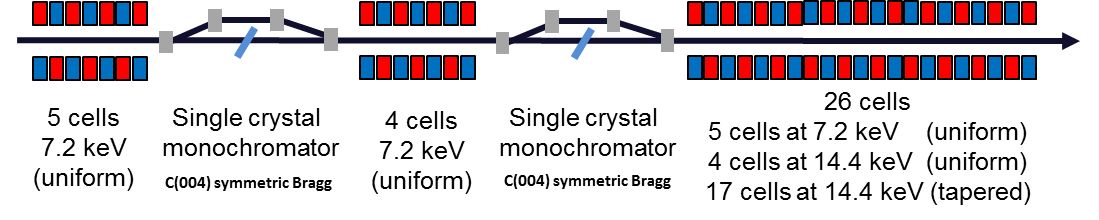}
\caption{Layout of the SASE2 undulator at the European XFEL configured for HXRSS and tapered operation, as discussed in this work.} \label{layoutund}
\end{figure}
HXRSS enables active spectral filtering by FEL amplification of a monochromatized SASE signal produced in the first part of an XFEL setup. Recently, a monochromatization scheme based on a single diamond crystal \cite{OURY5b} was successfully realized at the LCLS \cite{AMAN} and at SACLA \cite{INAG}.

The simplest realization of a HXRSS setup based on a single crystal monochromator is composed by three parts. In the first part, a SASE pulse is produced in the linear regime, as in conventional XFELs. In the second part the electron beam is sent through a bypass composed of a short (few-m long) magnetic chicane, which washes out the electron beam microbunching, enables a tunable delay with respect to the photon beam, and allows for the introduction of a monochromator on the radiation path. The monochromator itself is composed by a single, thin diamond crystal (in our case $100~\mu$m thin) in transmission geometry. The crystal works as a filter, imposing a narrow dip in the XFEL SASE spectrum around the Bragg energy. In the time domain, such dip corresponds to a `wake' of monochromatic radiation following the main XFEL pulse, which is transmitted unperturbed. When the relative delay between electrons and X-ray pulse is properly chosen, this wake can be used to seed the electron bunch and gets amplified up to saturation in the third part of the setup, composed by an output undulator. This procedure typically allows for a reduction in bandwidth down to almost the Fourier limit, and thus corresponds to a manyfold increase in the spectral density of the original X-ray pulse.

\begin{figure}
\includegraphics[width=0.5\textwidth]{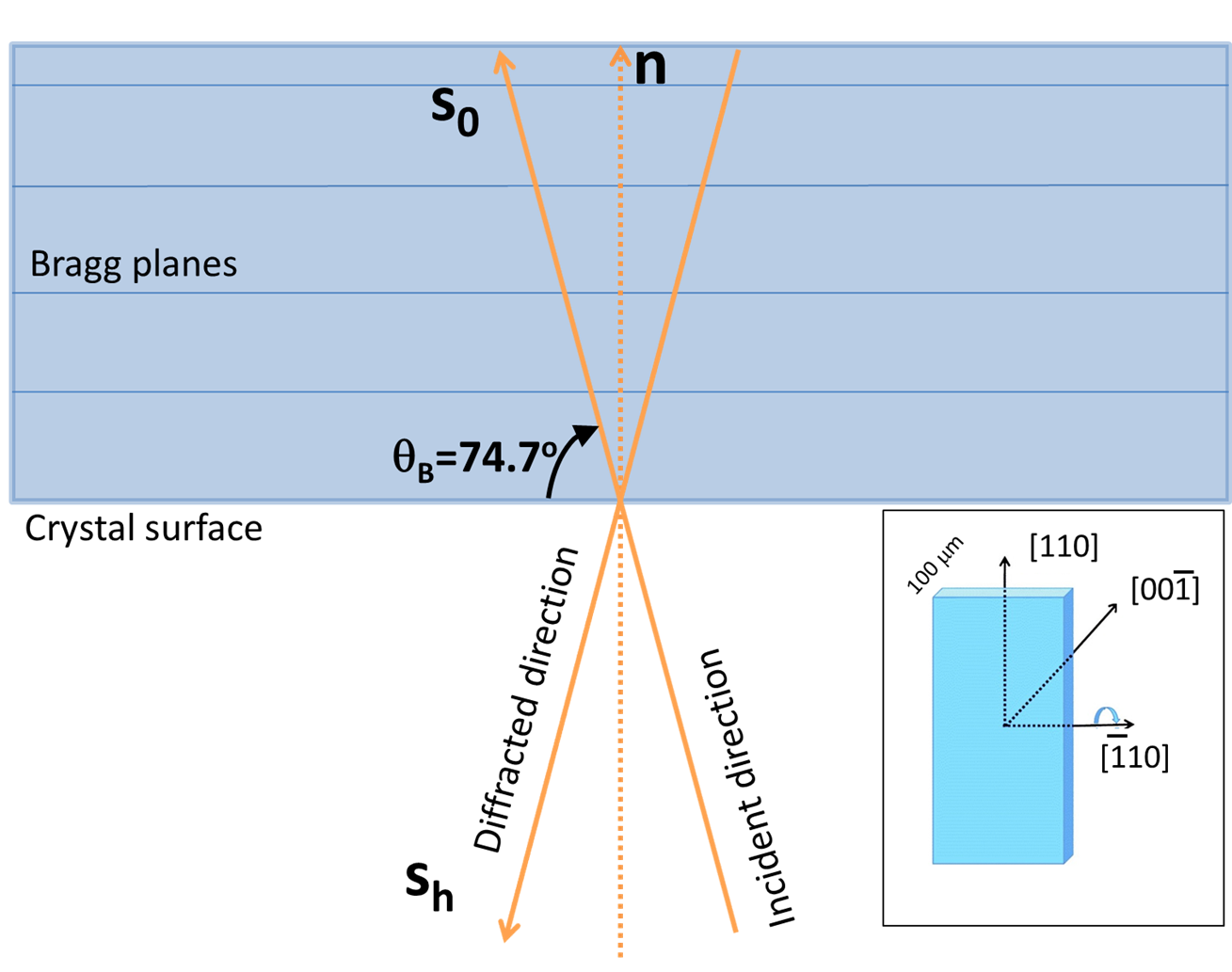}
\includegraphics[width=0.5\textwidth]{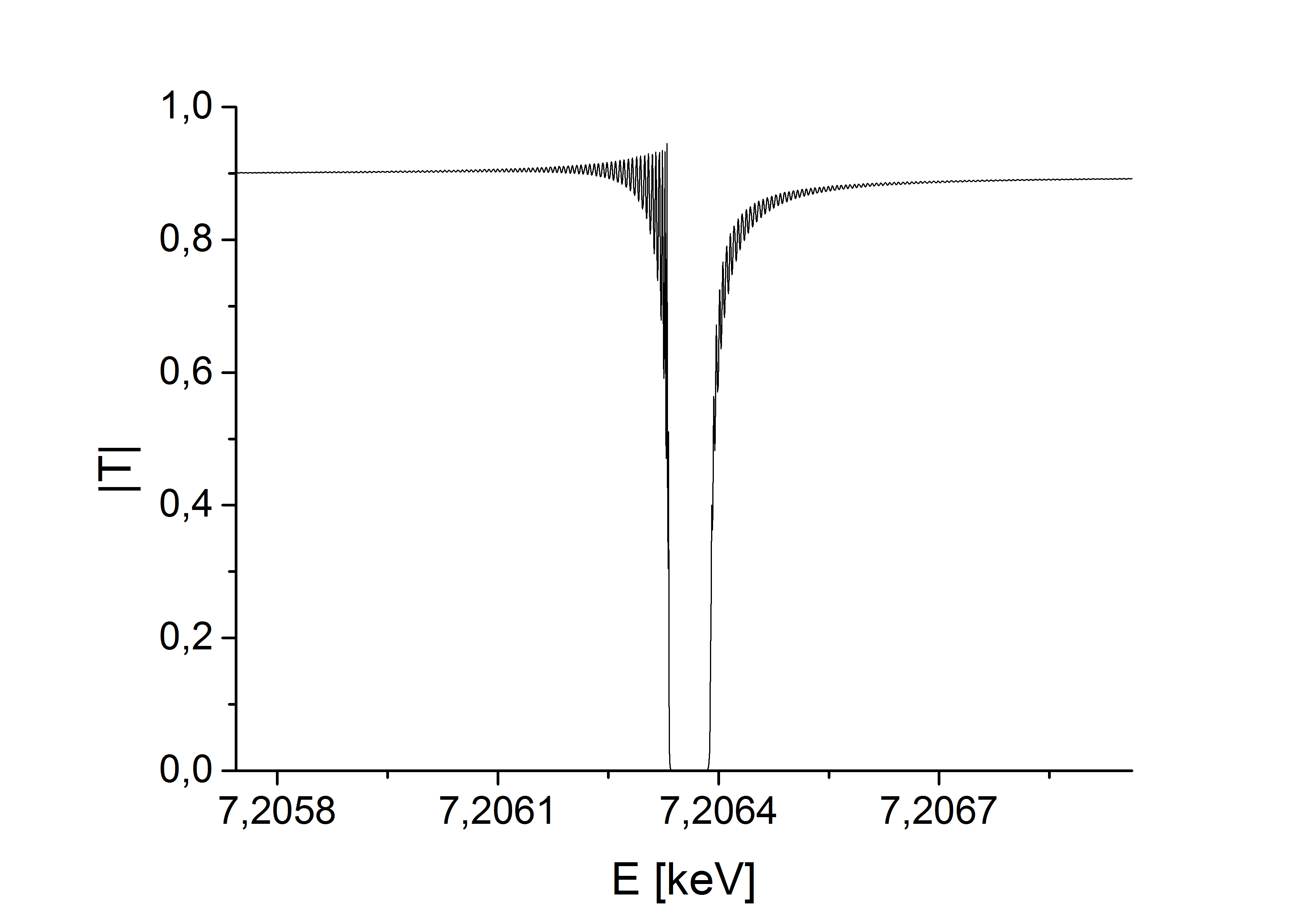}
\caption{Left: crystal geometry and C(004) symmetric Bragg reflection at $7.2$ keV. The inset shows the crystal cut, the pitch axis being $[\bar{1}10]$. Right: modulus of the transmission function for the C(004) symmetric bragg reflection at $7.2$ keV.} \label{crystal}
\end{figure}
When dealing with high-repetition rate XFELs, special care must be taken to avoid excessive heat-load of the diamond crystal. One way to increase the ratio between the seeded signal and the SASE noise is to perform the monochromatization process twice. At the position of the second crystal the signal is already almost Fourier-limited and monochromatization allows an increase of the seed signal, compared to the SASE background, of a factor roughly equal to the ratio between the SASE bandwidth and the seeded bandwidth, typically an order of magnitude. This increase in the signal-to-noise ratio can be used to diminish the length of the XFEL preceding the HXRSS setups, thus reducing the heat-load on the crystals \cite{OURY2}. This scheme, see Fig. \ref{layoutund} will be realized at the SASE2 line of the European XFEL \cite{XFELSEED}.

The HXRSS setup can be tuned in energy  by changing the pitch angle of the diamond crystal. The exploitation of several (symmetric and asymmetric) reflections allow in principle to cover all the XFEL spectrum starting from about $3$ keV. However, while increasing the energy (here we consider the remarkable energy point at $14.4$ keV) one faces the problem that the HXRSS efficiency decreases, due to a combination of several reasons. Suppose that one works with a given crystal and a fixed reflection, for example C(004), increasing the photon energy. This leads to a smaller and smaller Bragg angle. However, due to the presence of a spatio-temporal coupling phenomenon \cite{SHVI, ASYM}, the field transmitted by the crystal behaves as $E[t, x-ct \cot(\theta_B)]$. Therefore, when $\theta_B$ decreases, the superposition between electron beam and photon beam becomes worse and worse, leading to a decrease in efficiency. One may try to solve the issue by using other reflections, for C(444), where the Bragg's angle is very near to $\pi/2$ radians around the target energy of $14.4$ keV. However, as the Miller indexes increase the Darwin width narrows. As a result, the maximum of the impulse response of the filter decreases and therefore the seeded pulse become smaller, thus decreasing the overall efficiency. Furthermore, as the energy increases, the equivalent SASE shot-noise increases, as does the gain length. These factors concur to a  decrease in the final spectral density of self-seeded pulses at high energies, which we observed in simulations. Here we propose a scheme to overcome this problem, without changes to the HXRSS hardware.

The undulator system of the European XFEL is long enough to allow one to use CHG techniques to overcome this problem without changing the HXRSS hardware, e.g. without the need for a crystal optimized for high-energy applications. Near saturation, the bunching at harmonics of the fundamental begins to increase, the bunching being driven by the interaction of the first harmonic and the magnetic field of the undulator with the electron beam. This mechanism is well-known goes under the name of Coherent Harmonic Generation (CHG) and was extensively studied both theoretically and experimentally \cite{SCHM}-\cite{SPEZ}. To see how we can take advantage of CHG, let us consider the $Fe^{57}$ nuclear resonance at $14.4$ keV as target energy. With reference to Fig. \ref{layoutund} one seeds at $7.2$ keV, using the C(004) symmetric reflection, where HXRSS is quite efficient, Fig. \ref{crystal}. Following the two HXRSS setups one lets the X-ray beam at $7.2$ keV to be amplified in the first part of the third undulator.  In order to exploit the CHG mechanism we now tune the fundamental of the second part of the radiator at the second harmonic, which is the target energy $14.4$ keV. Simulation shows that the second harmonic content in the electron beam bunching is large enough to reach saturation in a few segments. The rest can be used to increase the output power of about an order of magnitude. This tenfold increase, together with a likewise narrowing of the spectral bandwidth yields an advantage of two orders of magnitude in the spectral density of the output radiation compared to SASE.

In particular, our simulations shows that one can indeed reach about $4 \cdot 10^{13}$ ph/s/meV in the case of seeded-tapered pulses at the SASE2 line of the European XFEL at $14.4$ keV. These simulations studies are reported in a detailed way in the next section \ref{fels}, while in section \ref{sec:conc} we come to conclusions.

\section{\label{fels} FEL studies}

\begin{figure}[tb]
\includegraphics[width=0.5\textwidth]{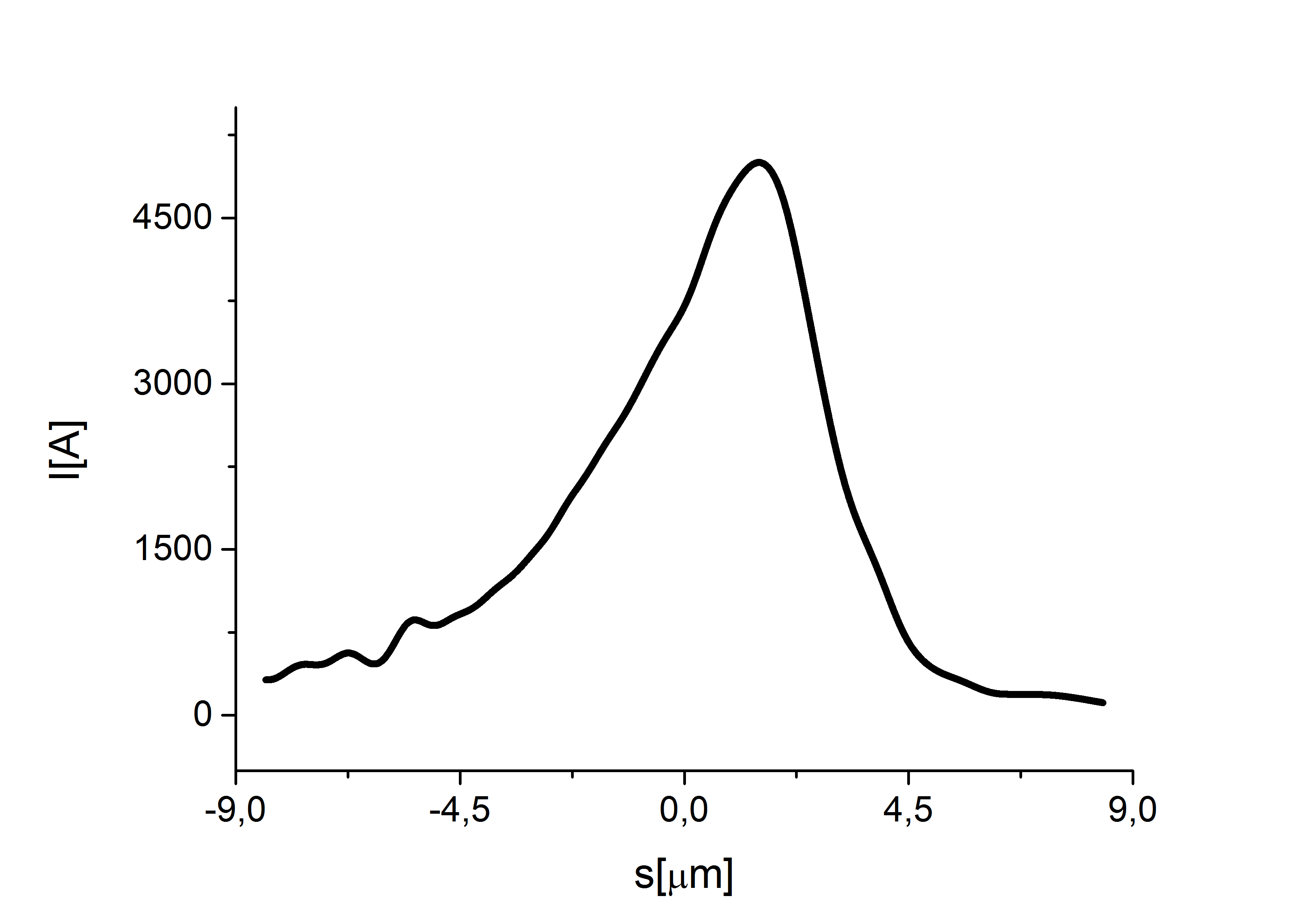}
\includegraphics[width=0.5\textwidth]{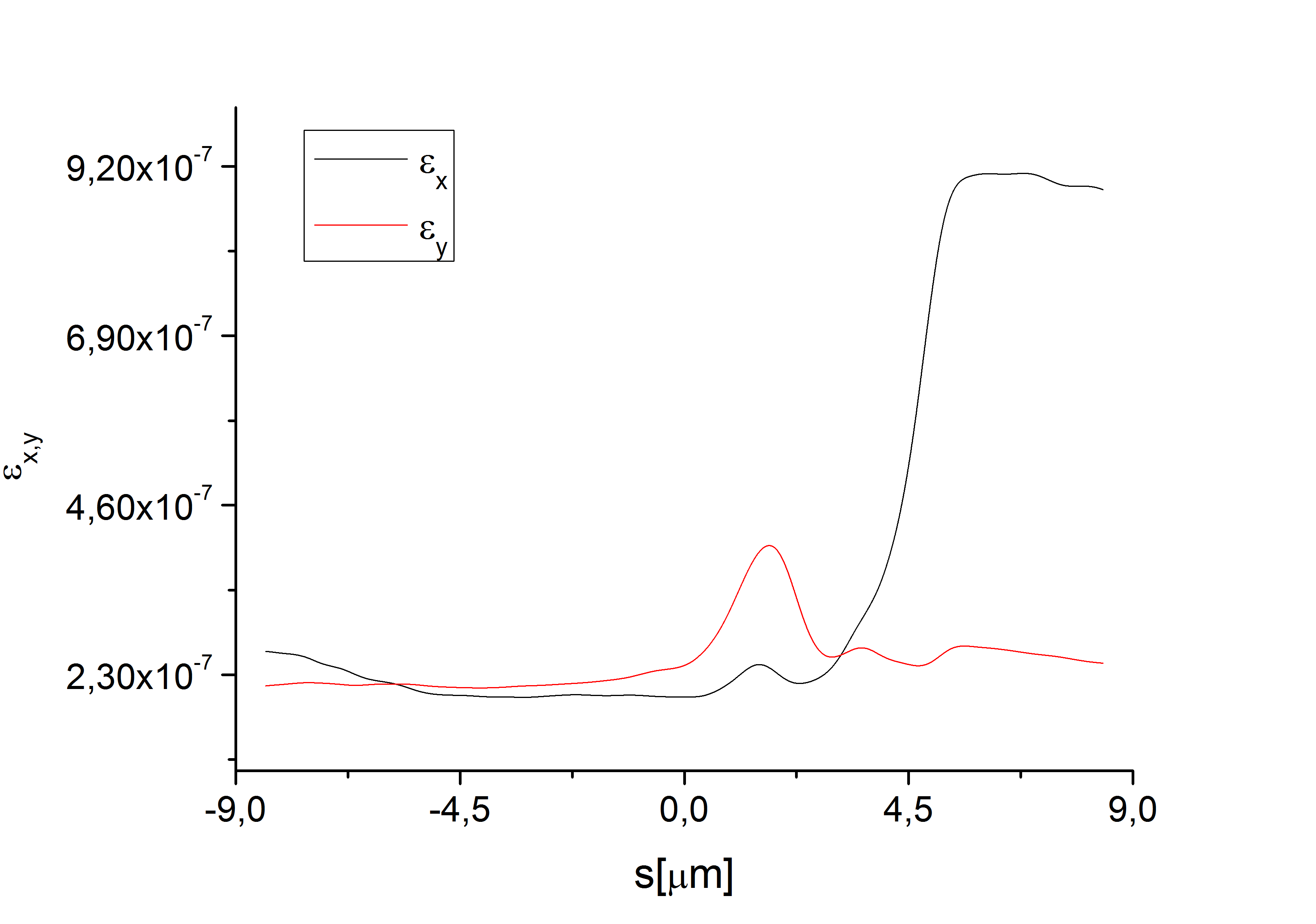}
\includegraphics[width=0.5\textwidth]{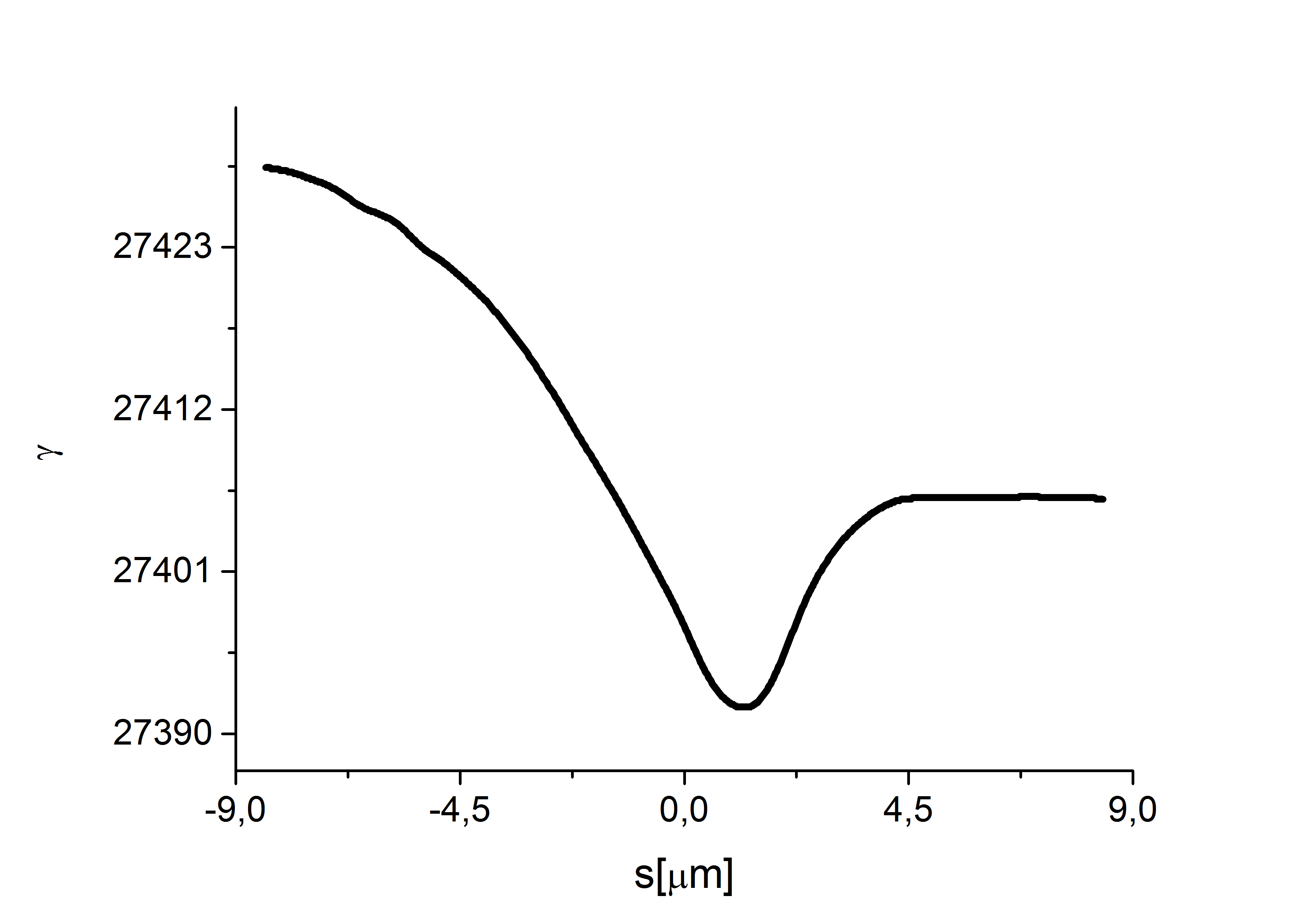}
\includegraphics[width=0.5\textwidth]{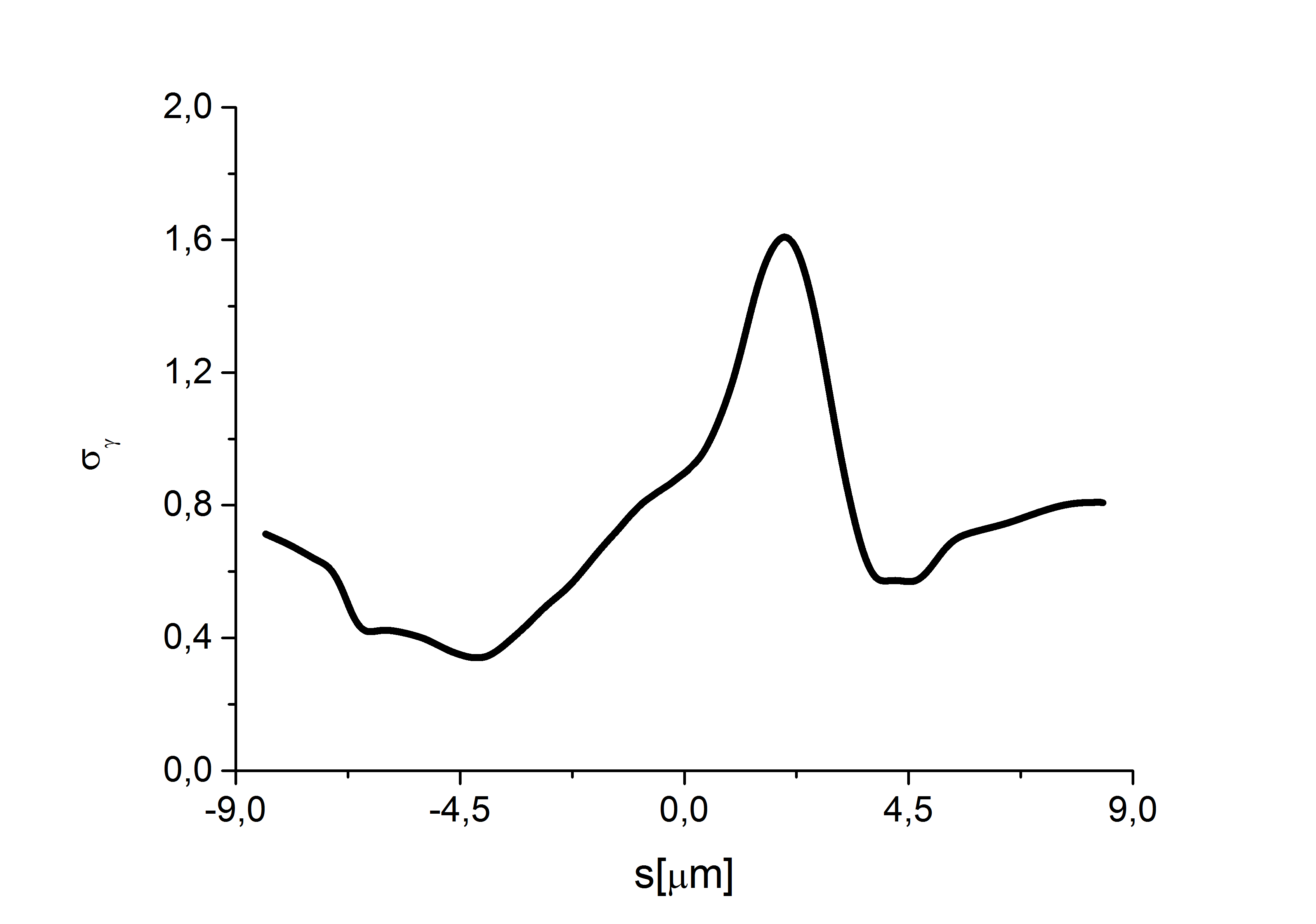}
\begin{center}
\includegraphics[width=0.5\textwidth]{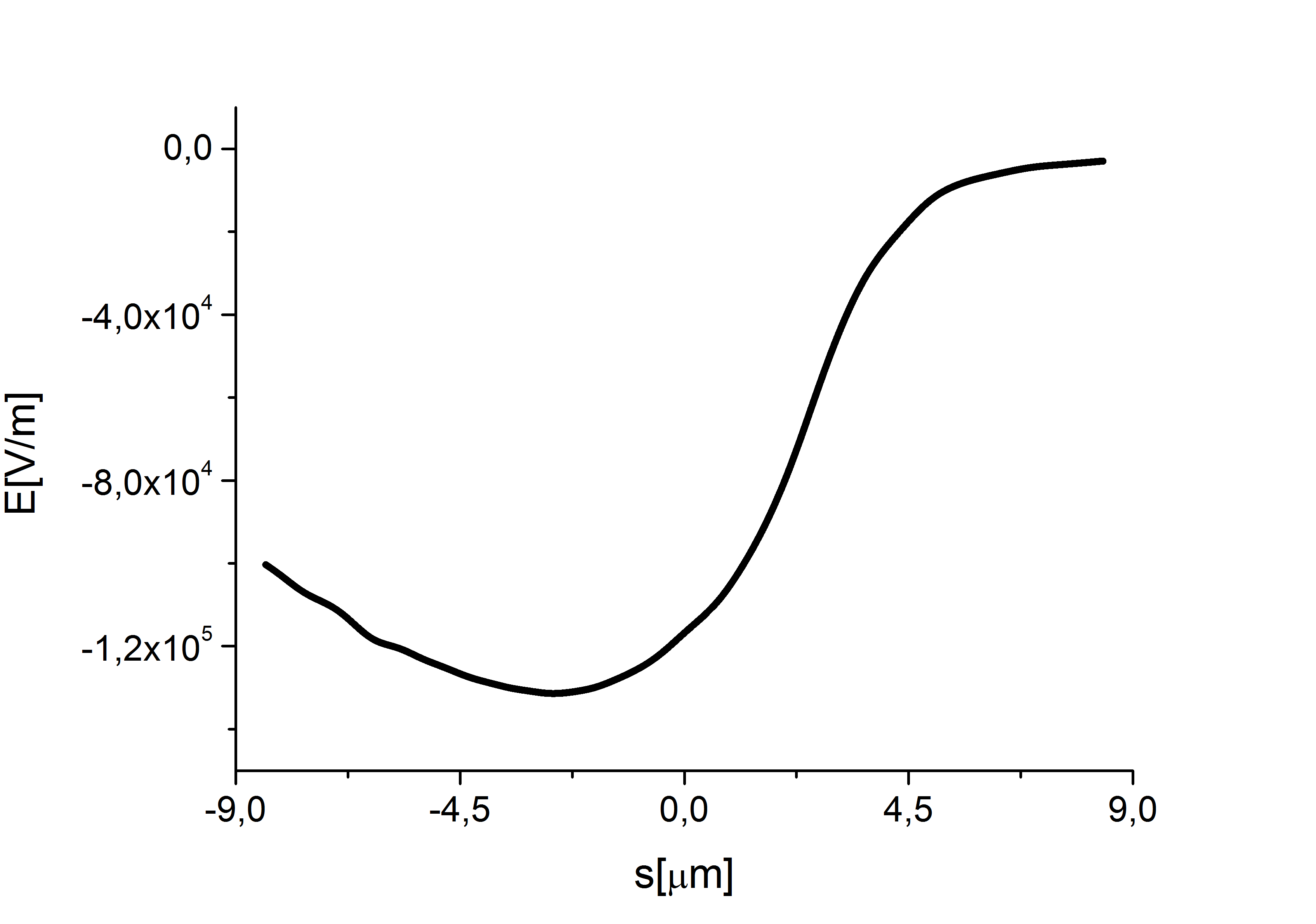}
\end{center}
\caption{Results from electron beam start-to-end simulations at the entrance of the SASE2 line \cite{S2ER} for the hard X-ray case for the 17.5 GeV mode of operation. First Row, Left: current profile. First Row, Right: normalized emittance as a function of the position inside the electron beam. Second Row, Left: energy profile along the beam in units of the relativistic factor $\gamma$. Second Row, right: rms energy spread profile in units of the relativistic factor $\gamma$ along the beam. Bottom row: resistive wakefields in the SASE2 undulator.} \label{s2e}
\end{figure}

The source discussed in the previous section, Fig. \ref{layoutund} was studied using the code Genesis \cite{GENE}. We ran a statistical analysis consisting of $100$ runs.  Start-to-end simulations \cite{S2ER} were used as input information for GENESIS, and define the electron beam quality, see Fig. \ref{s2e}. The main parameters summarizing the mode of operation of the European XFEL discussed here can be found in Table \ref{tt1}.

\begin{table}
\caption{Main parameters for the mode of operation of SASE2 considered here.}

\begin{small}\begin{tabular}{ l c c}
\hline & ~ Units &  ~ \\ \hline
Undulator period      & mm                  & 40     \\
Periods per cell      & -                   & 125   \\
Total number of cells & -                   & 35    \\
Intersection length   & m                   & 1.1   \\
Energy                & GeV                 & 17.5    \\
Charge                & pC                  & 100\\
\hline
\end{tabular}\end{small}
\label{tt1}
\end{table}

\begin{figure}
\includegraphics[width=0.50\textwidth]{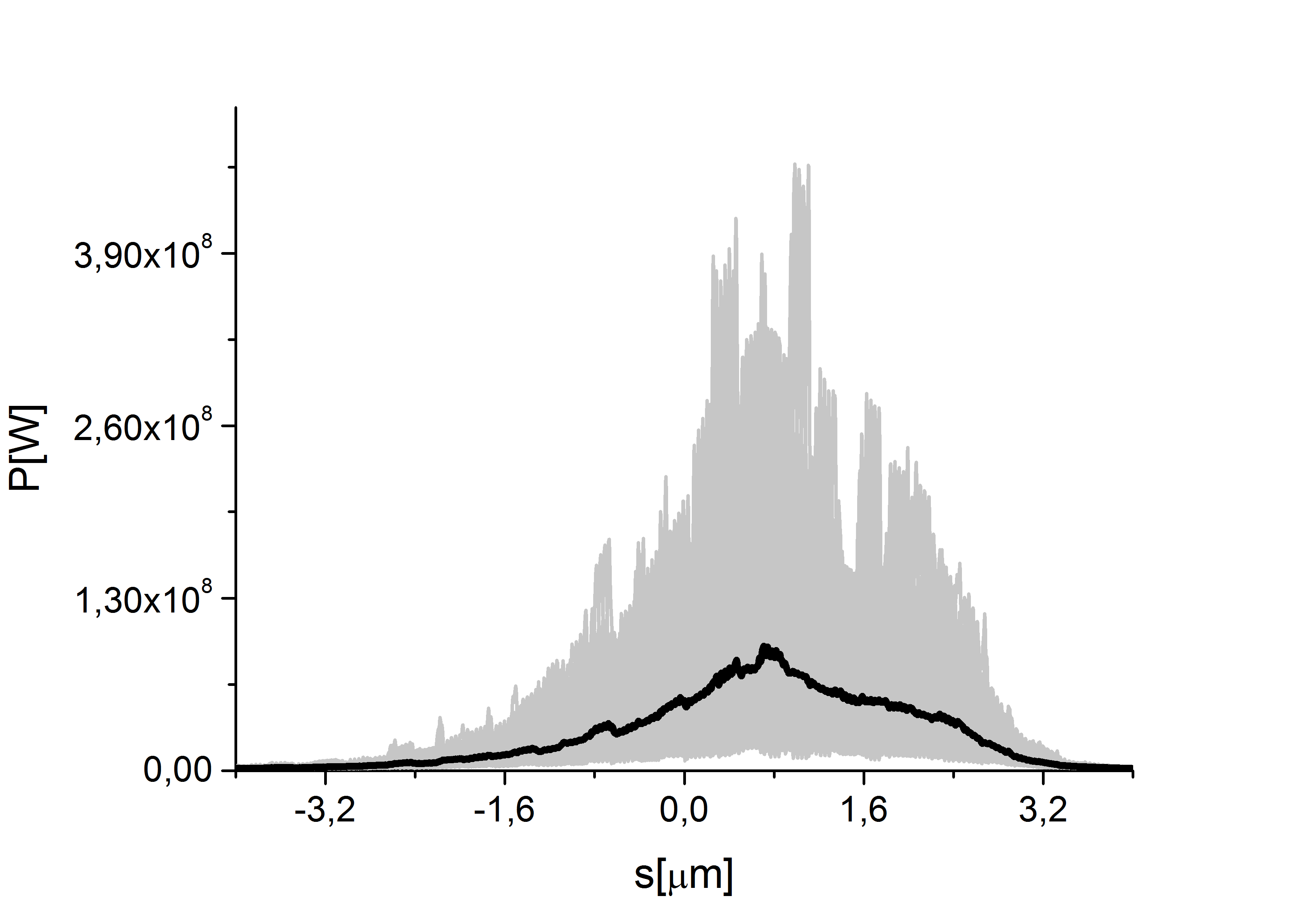}
\includegraphics[width=0.50\textwidth]{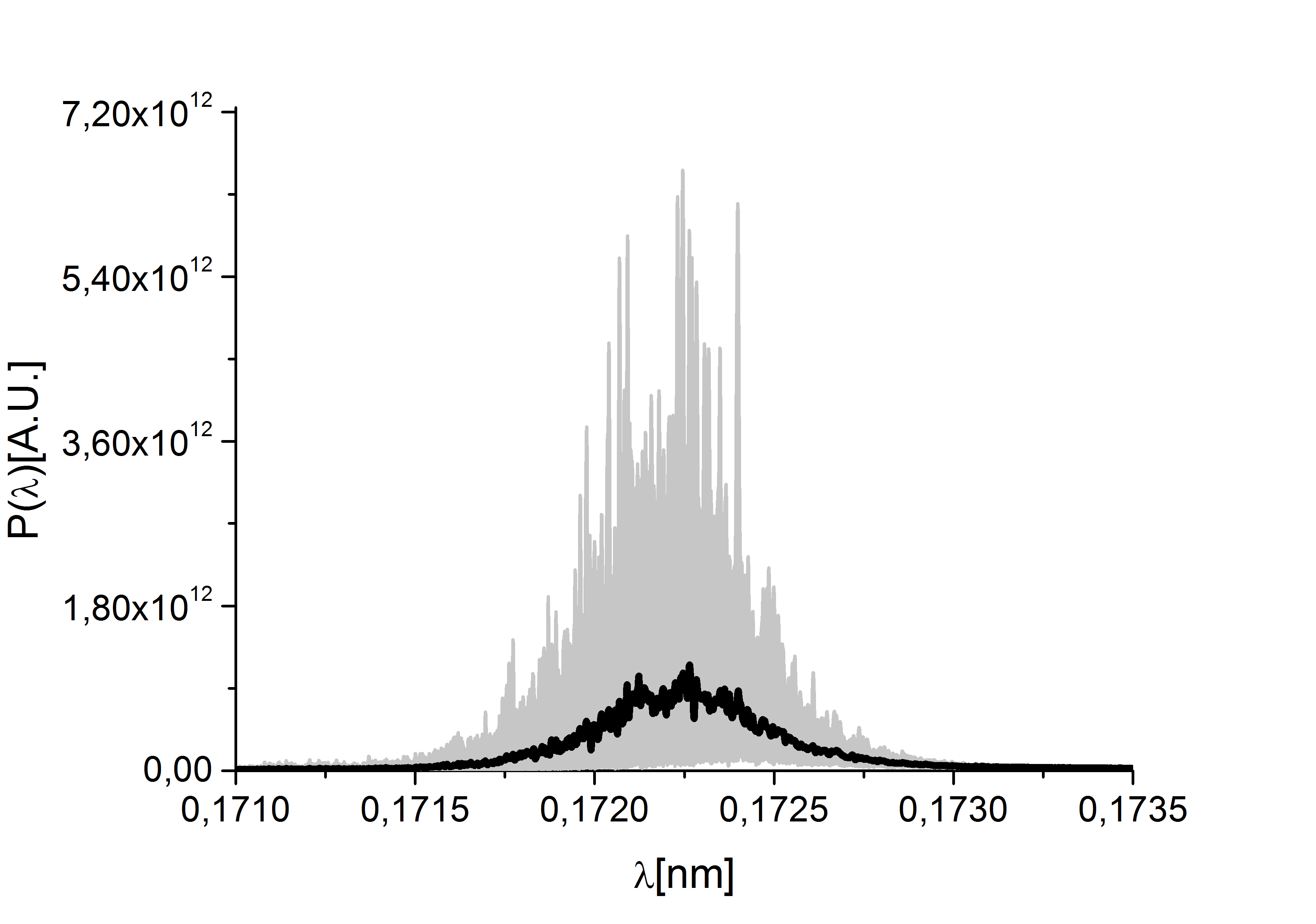}
\caption{Power distribution and spectrum of the SASE x-ray pulse at the exit of the first undulator (5 segments). Grey lines refer to single shot realizations, the black
line refers to the average over a hundred realizations.} \label{psp1in}
\end{figure}
The first five undulator segments produce SASE radiation at $7.2$ keV, one half of the target energy. The output power and spectrum are shown in Fig. \ref{psp1in}. When considering application of HXRSS methods to the European XFEL, one always needs to account the high-repetition rate of the setup, which poses strict limits on the energy per pulse incident onto the crystal. This can be evaluated by calculating the area under the power distribution in Fig. \ref{psp1in}, and gives an average of about $0.75~\mu$J per pulse, which is below the limiting heat load level even for the full-repetition rate of the European XFEL case.

\begin{figure}
\includegraphics[width=0.50\textwidth]{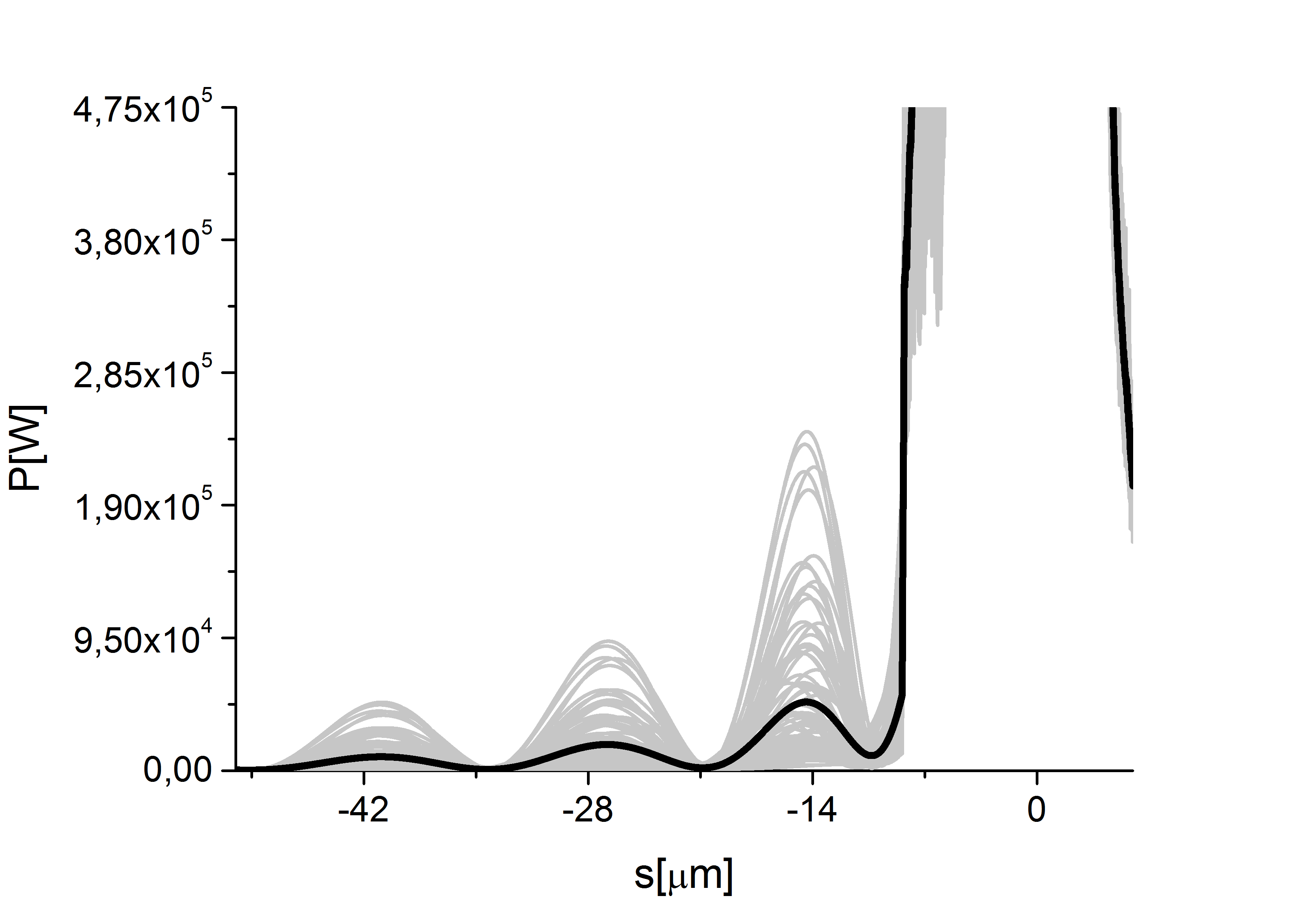}
\includegraphics[width=0.50\textwidth]{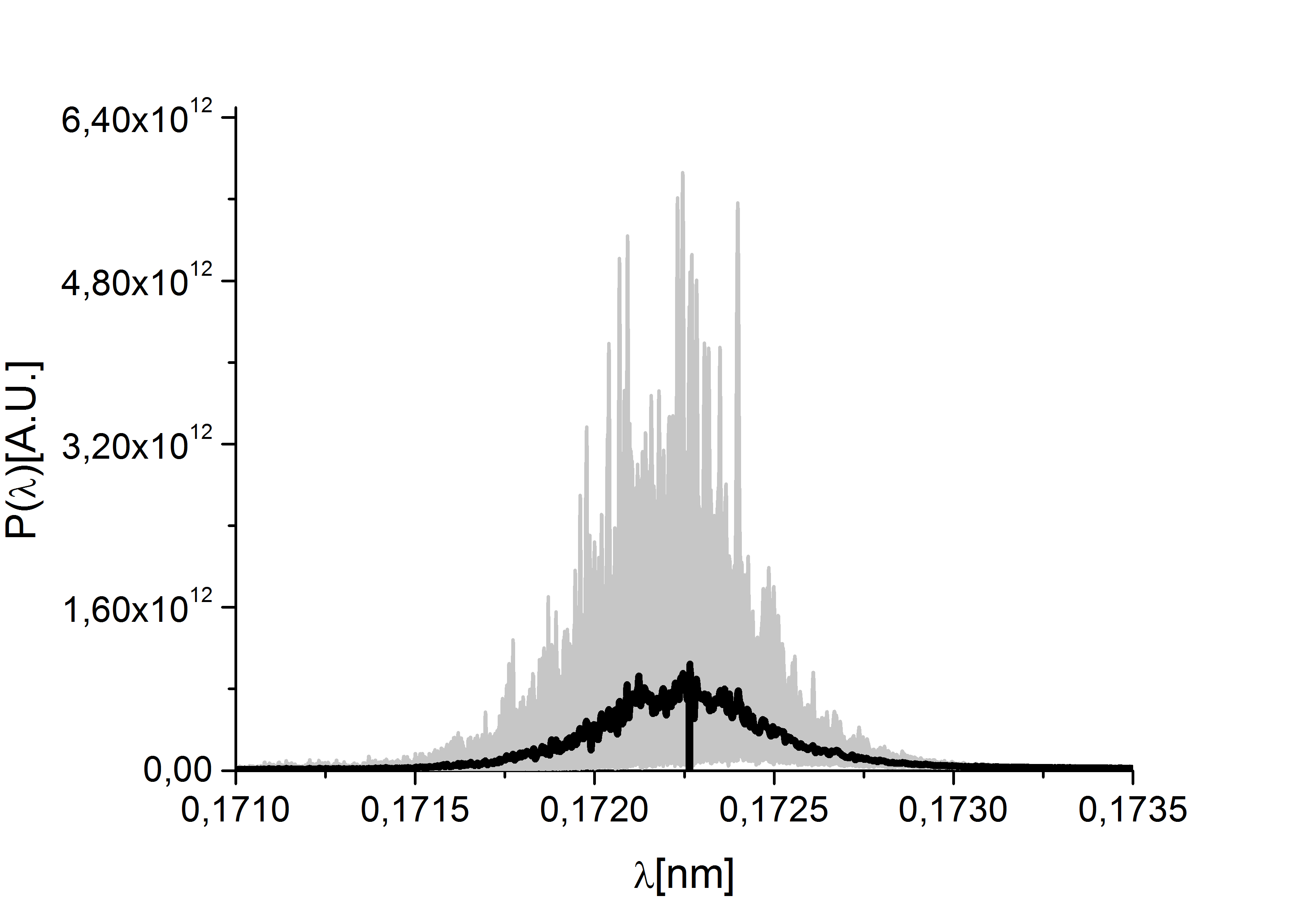}
\caption{Power distribution and spectrum after the first HXRSS monochromator.} \label{psp1out}
\end{figure}
Fig. \ref{psp1out} shows the outcome of the filtering process. Following the first crystal, the X-ray seed pulse proceeds through the second undulator in Fig. \ref{layoutund}, where it is amplified by the interaction with the electron beam.

\begin{figure}
\includegraphics[width=0.50\textwidth]{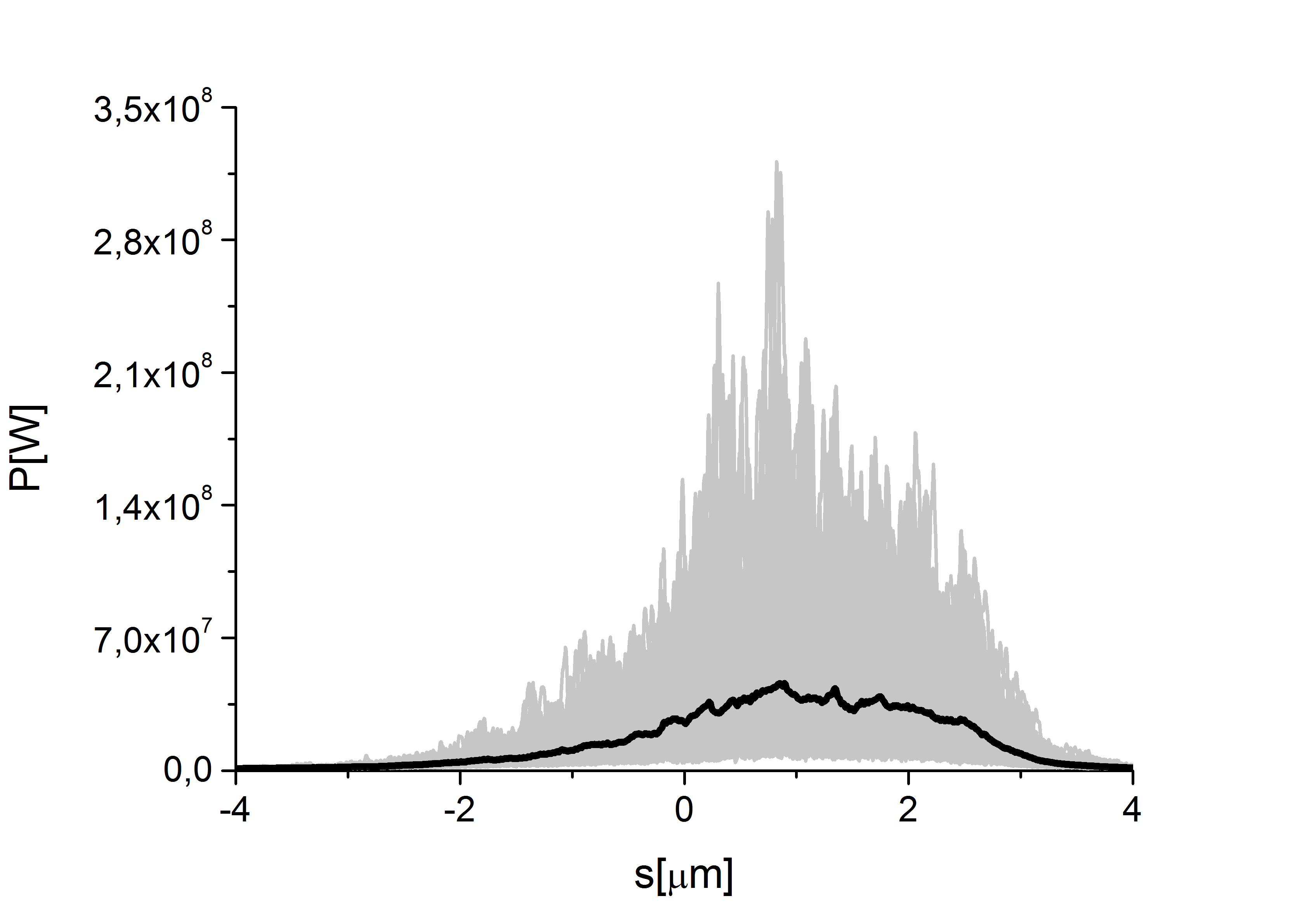}
\includegraphics[width=0.50\textwidth]{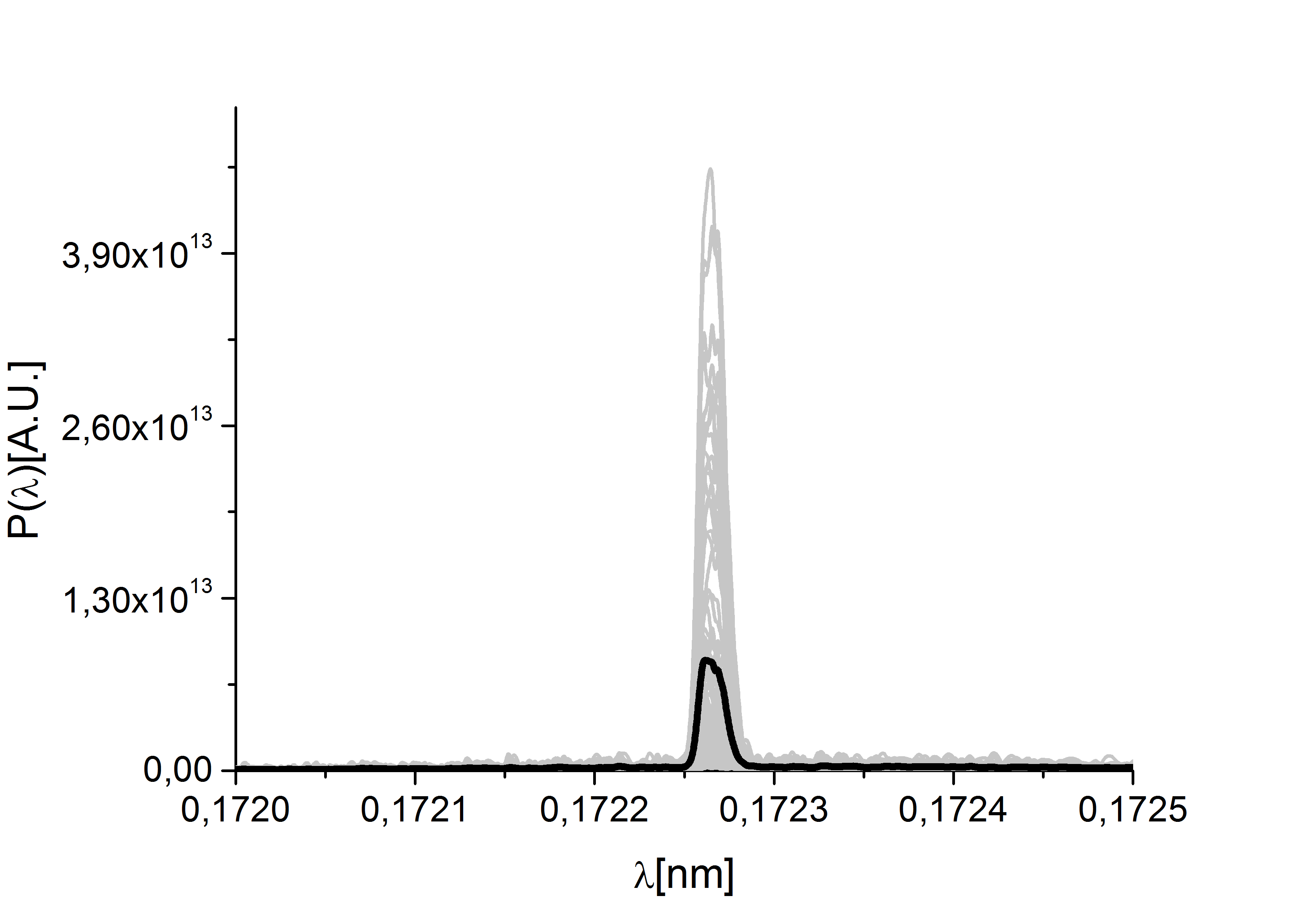}
\caption{Power distribution and spectrum of the SASE x-ray pulse at the exit of the second undulator (4 segments). Grey lines refer to single shot realizations, the black
line refers to the average over a hundred realizations.} \label{psp2in}
\end{figure}
\begin{figure}
\includegraphics[width=0.50\textwidth]{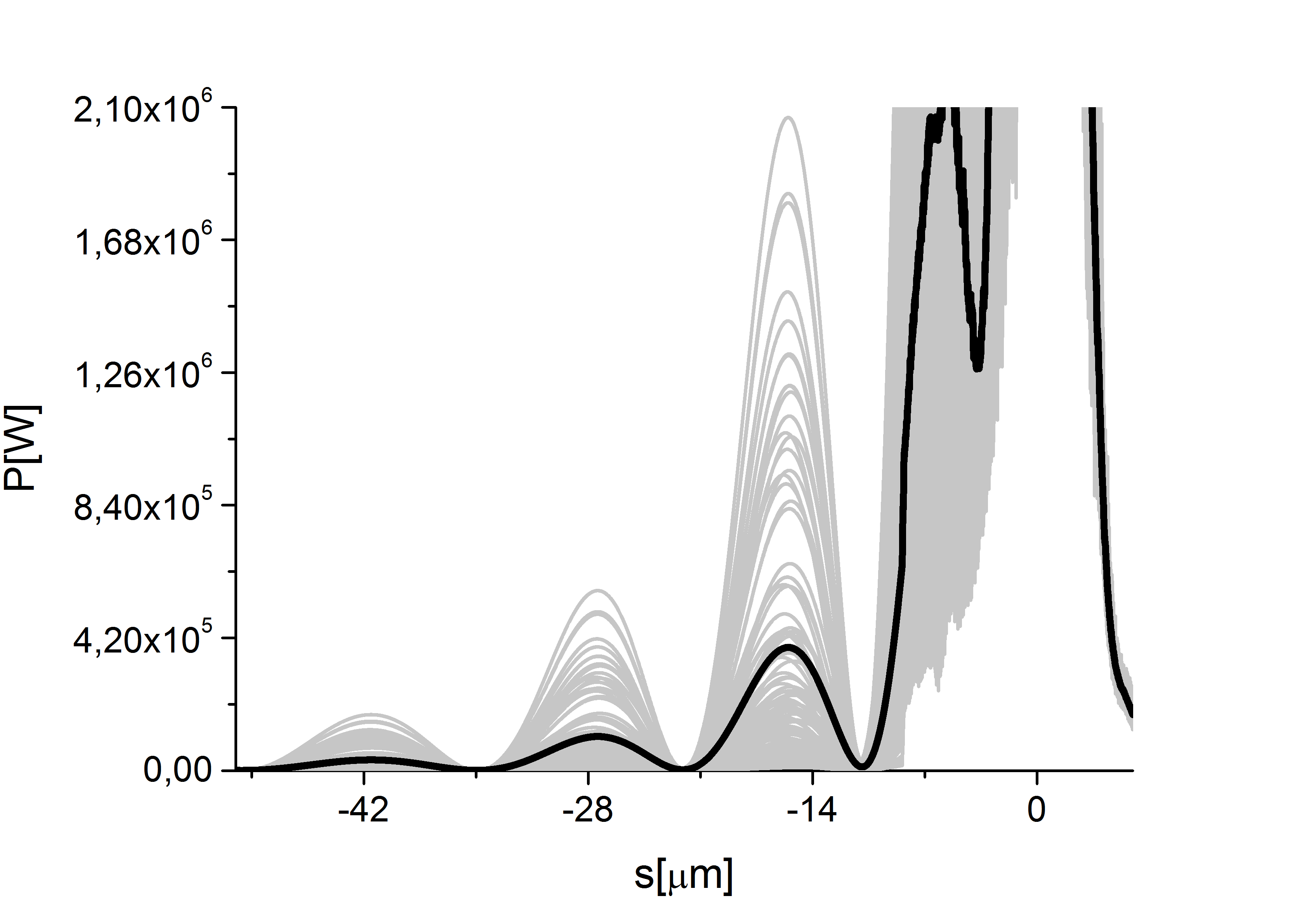}
\includegraphics[width=0.50\textwidth]{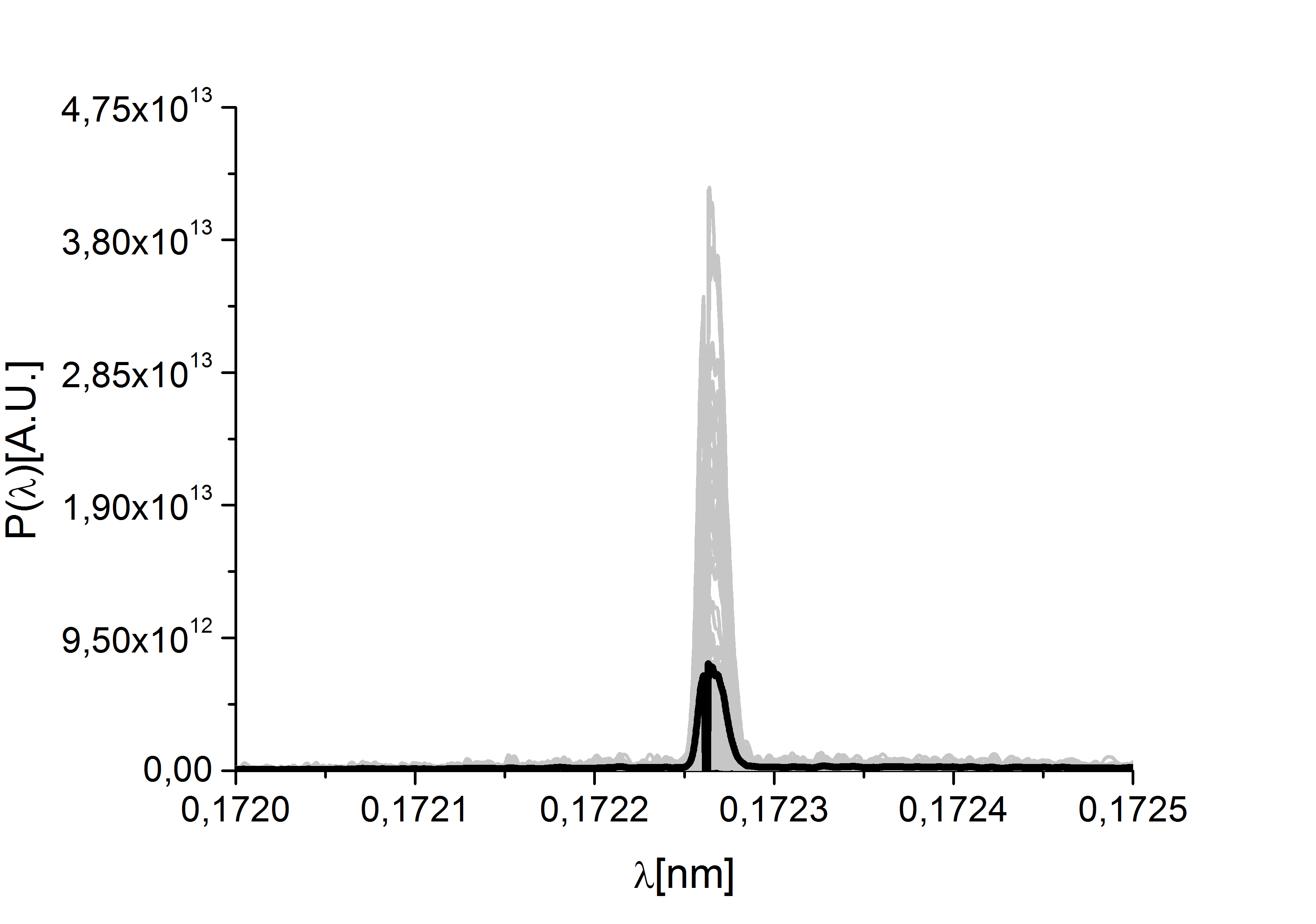}
\caption{Power distribution and spectrum after the second HXRSS monochromator. Grey lines refer to single shot realizations, the black
line refers to the average over a hundred realizations.} \label{psp2out}
\end{figure}

Fig. \ref{psp2in} illustrates power and spectrum at the exit of the second undulator. The relatively large spectral background in Fig. \ref{psp2in} (right) is due to the competition with the SASE process, and is visible due to the low seed power level from the first undulator. Note that the power level in the second pulse is lower than that on the first, and the average energy per pulse incident on the second crystal is about $0.44~\mu$J. However, the main peak in Fig. \ref{psp2in} (right) shows that at the seed frequency the spectral density is much higher (about one order of magnitude) than that at the first crystal, Fig. \ref{psp1in} (right). As a result, also the seed level after the second crystal, Fig. \ref{psp2out} (left) is about an order of magnitude larger than that after the first crystal, Fig. \ref{psp1out} (left). This is the advantage of the two-chicane scheme, which betters the signal-to-noise ratio (signal being the seeded FEL, noise being the SASE FEL pulse) of a factor roughly equal to the ratio between the SASE FEL bandwidth and the seeded FEL bandwidth, roughly equal to an order of magnitude.

After the second crystal, the seed signal is amplified into the first four segments of the final in Fig. \ref{layoutund}. At this point, the electron beam is significantly bunched at the second harmonic of the fundamental, around $14.4$ keV, and the microbunching is well correlated along the longitudinal direction, because it is driven by the fundamental. Hence the idea, first considered but not fully exploited in \cite{OURYFF} to tune the final part of the radiator at $14.4$ keV in order to exploit such bunching.

\begin{figure}
\begin{center}
\includegraphics[width=0.50\textwidth]{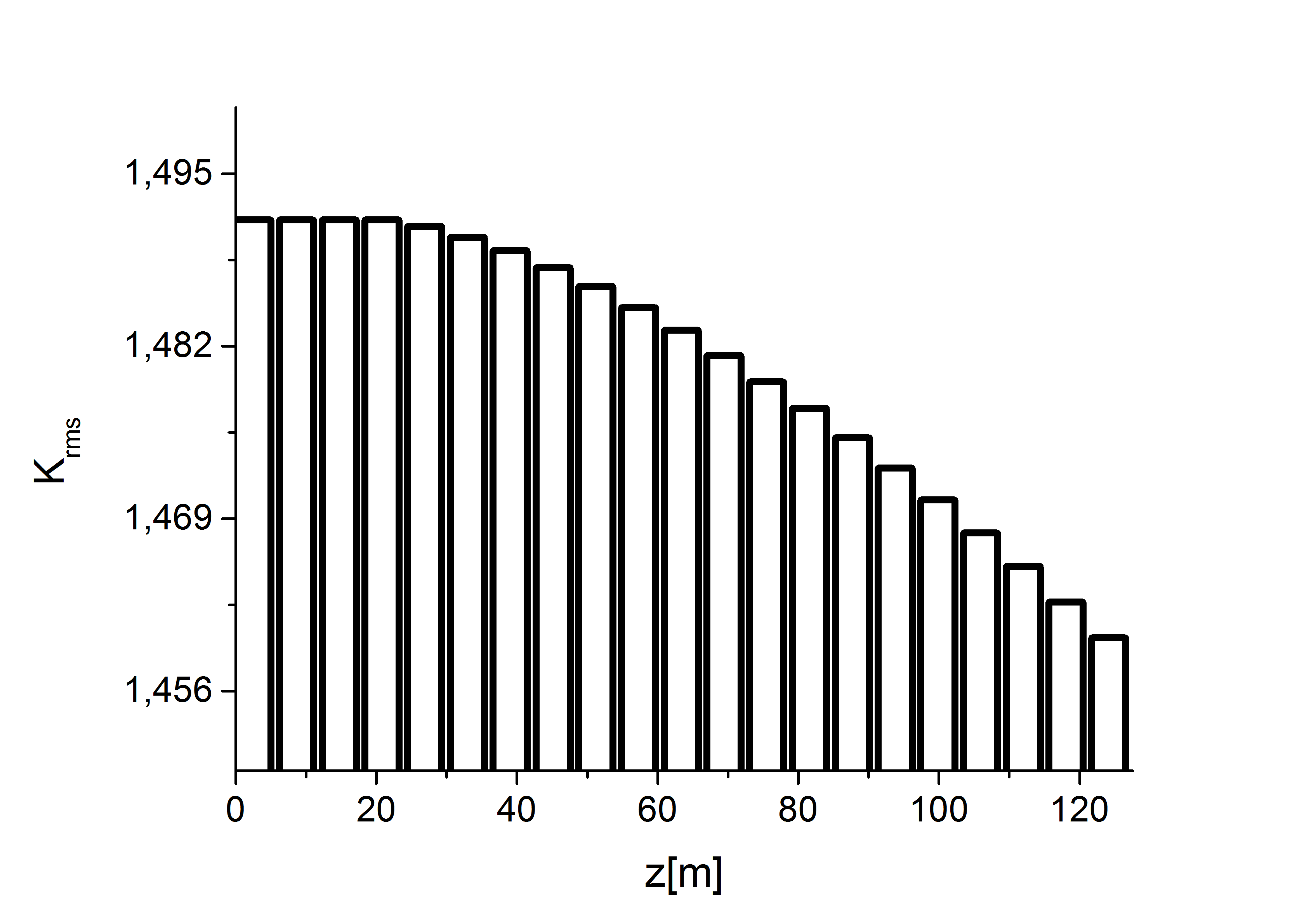}
\end{center}
\caption{Taper configuration for the output undulator (21 segments). } \label{Taplaw}
\end{figure}

The last $17$ cells are tapered segment by segment as illustrated in Fig. \ref{Taplaw}. The optimal tapering was found on an empirical basis, in order to optimize the final spectral density of the output signal.

\begin{figure}
\includegraphics[width=0.50\textwidth]{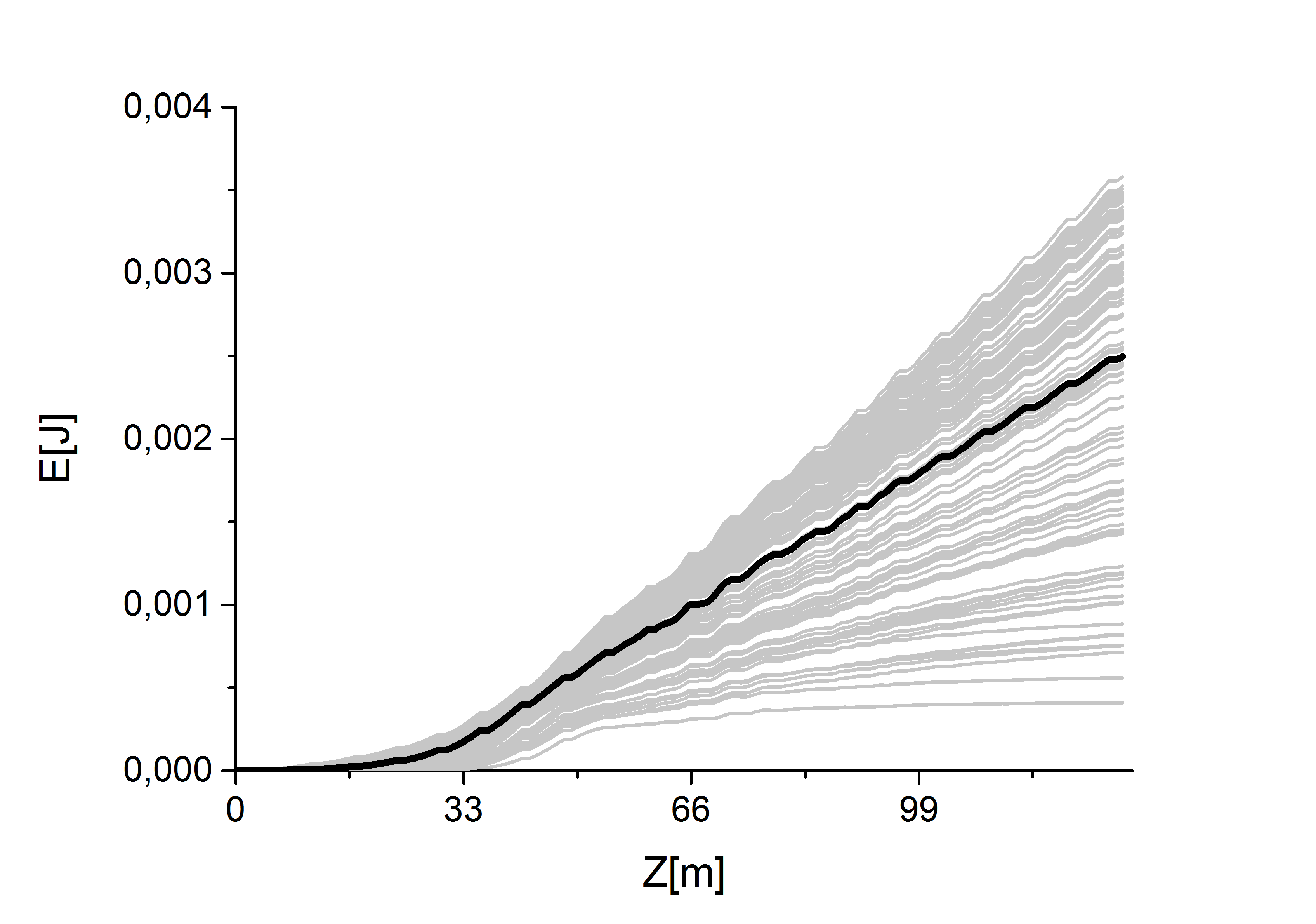}
\includegraphics[width=0.50\textwidth]{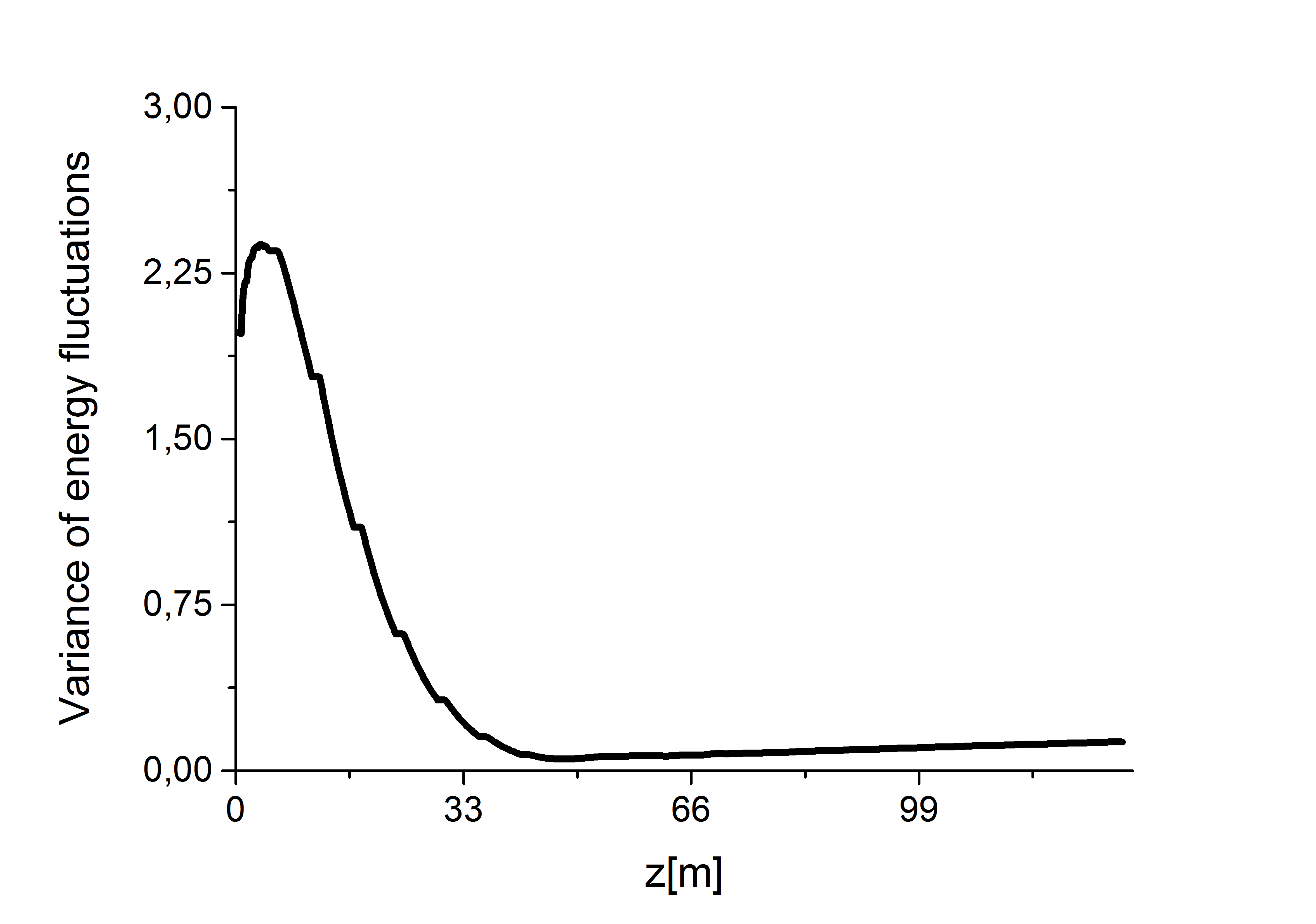}
\caption{Energy and variance of energy fluctuations of the seeded FEL pulse as a function of the
distance inside the output undulator. Grey lines refer to single shot realizations, the black
line refers to the average over a hundred realizations.} \label{enevar}
\end{figure}
In Fig. \ref{enevar} we plot the energy and variance of energy fluctuations of the seeded FEL pulse as the latter develops along the undulator. Note the high maximum variance in Fig. \ref{enevar} (right), which can be ascribed to the statistical properties of the second harmonic. From Fig. \ref{enevar} (left) it can be seen that we are able to produce pulses up to a few mJ energy.

\begin{figure}
\includegraphics[width=0.50\textwidth]{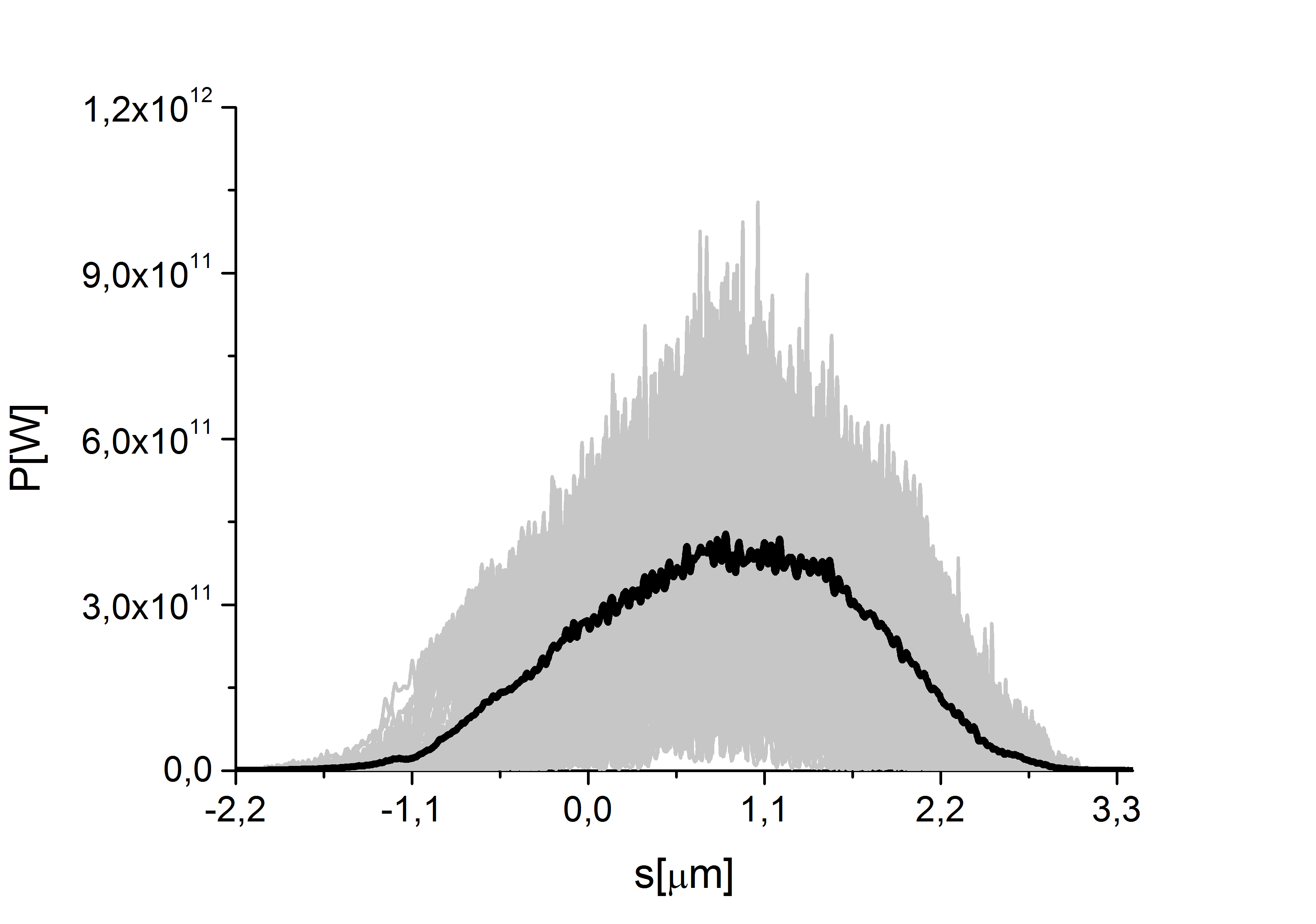}
\includegraphics[width=0.50\textwidth]{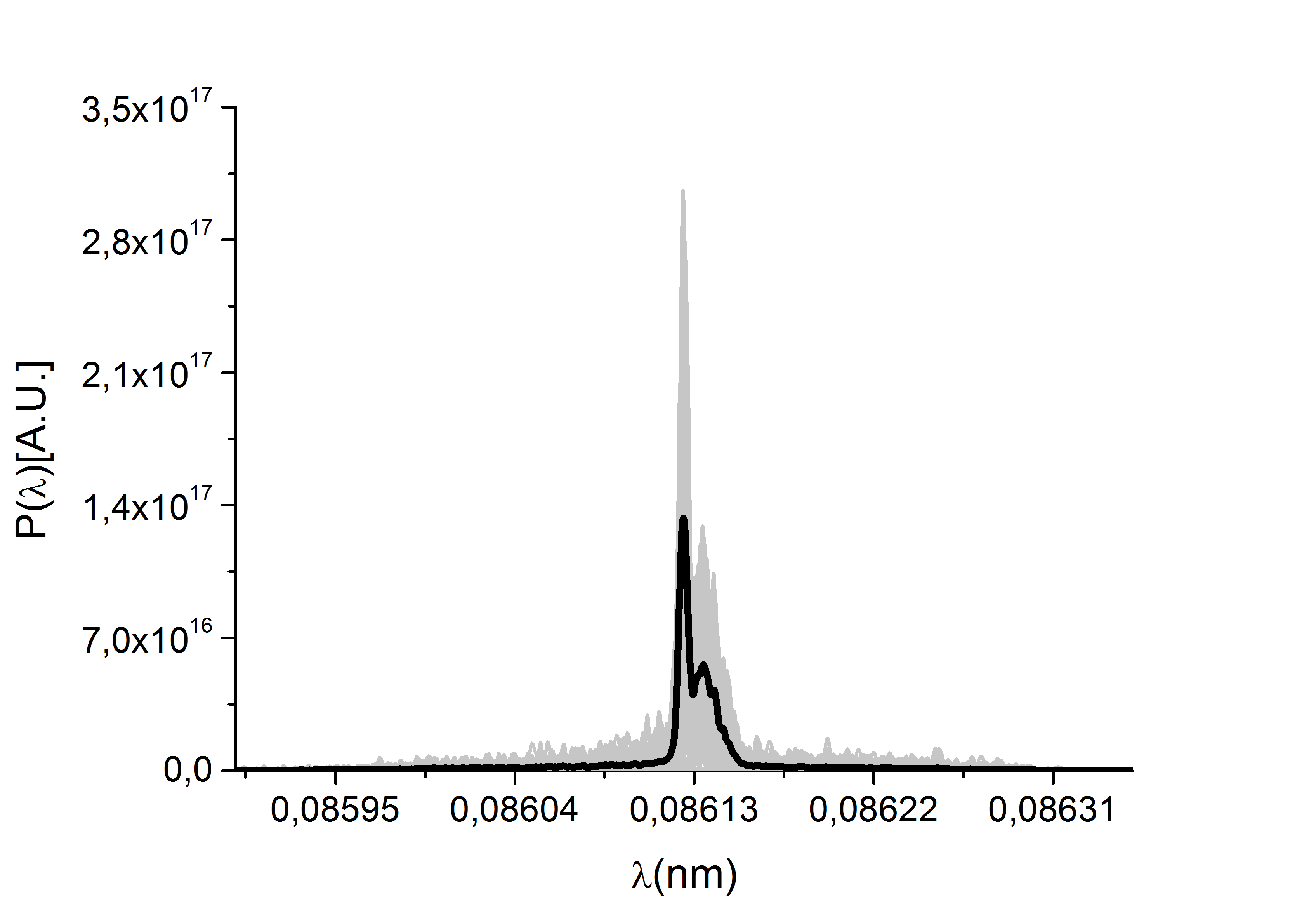}
\caption{Power distribution and spectrum at the exit of the setup. Grey lines refer to single shot realizations, the black line refers to the average over a hundred realizations.} \label{pspouttap}
\end{figure}
In Fig. \ref{pspouttap} one can see the final output of our setup in terms of power and spectrum.

\begin{figure}
\includegraphics[width=0.50\textwidth]{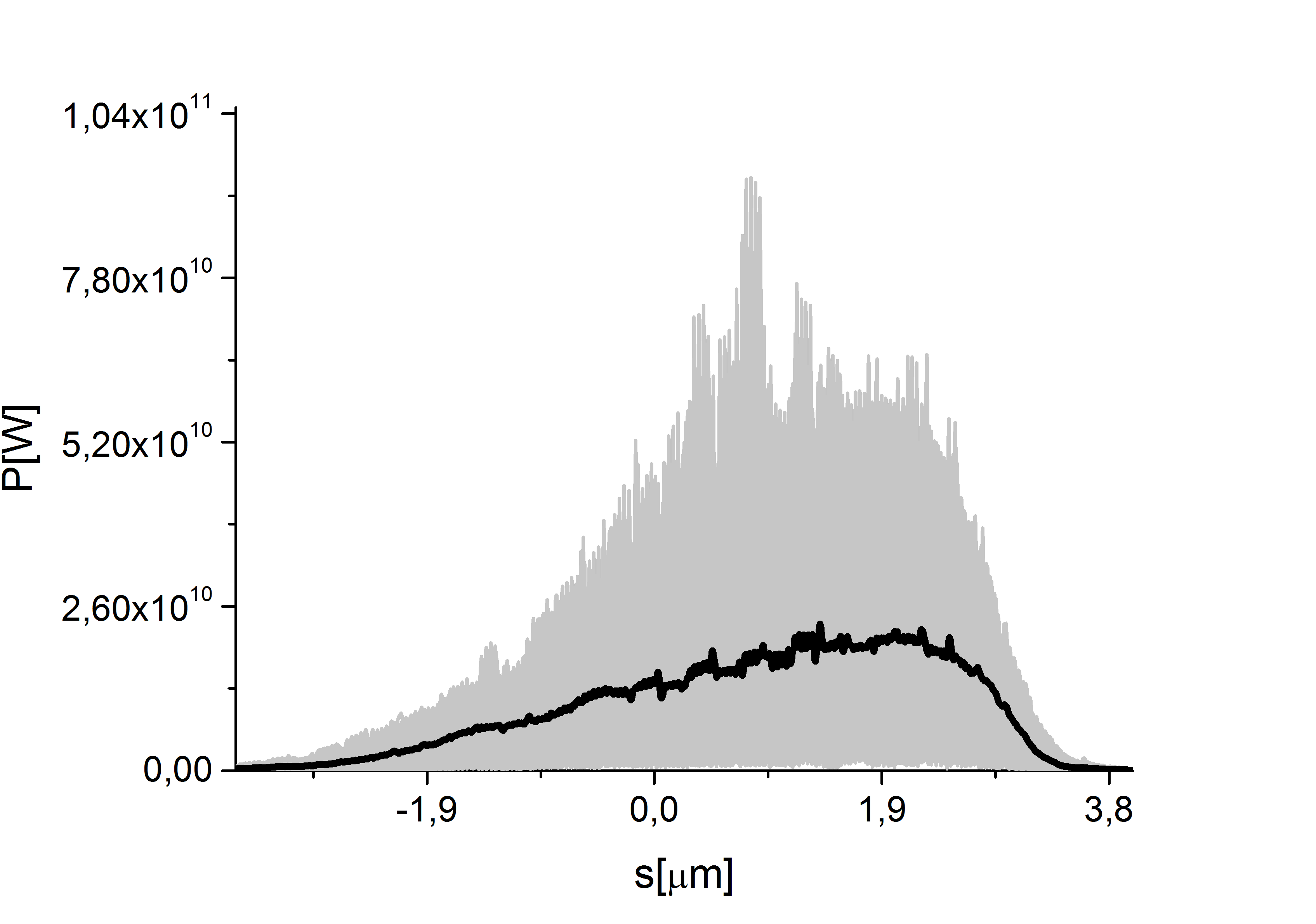}
\includegraphics[width=0.50\textwidth]{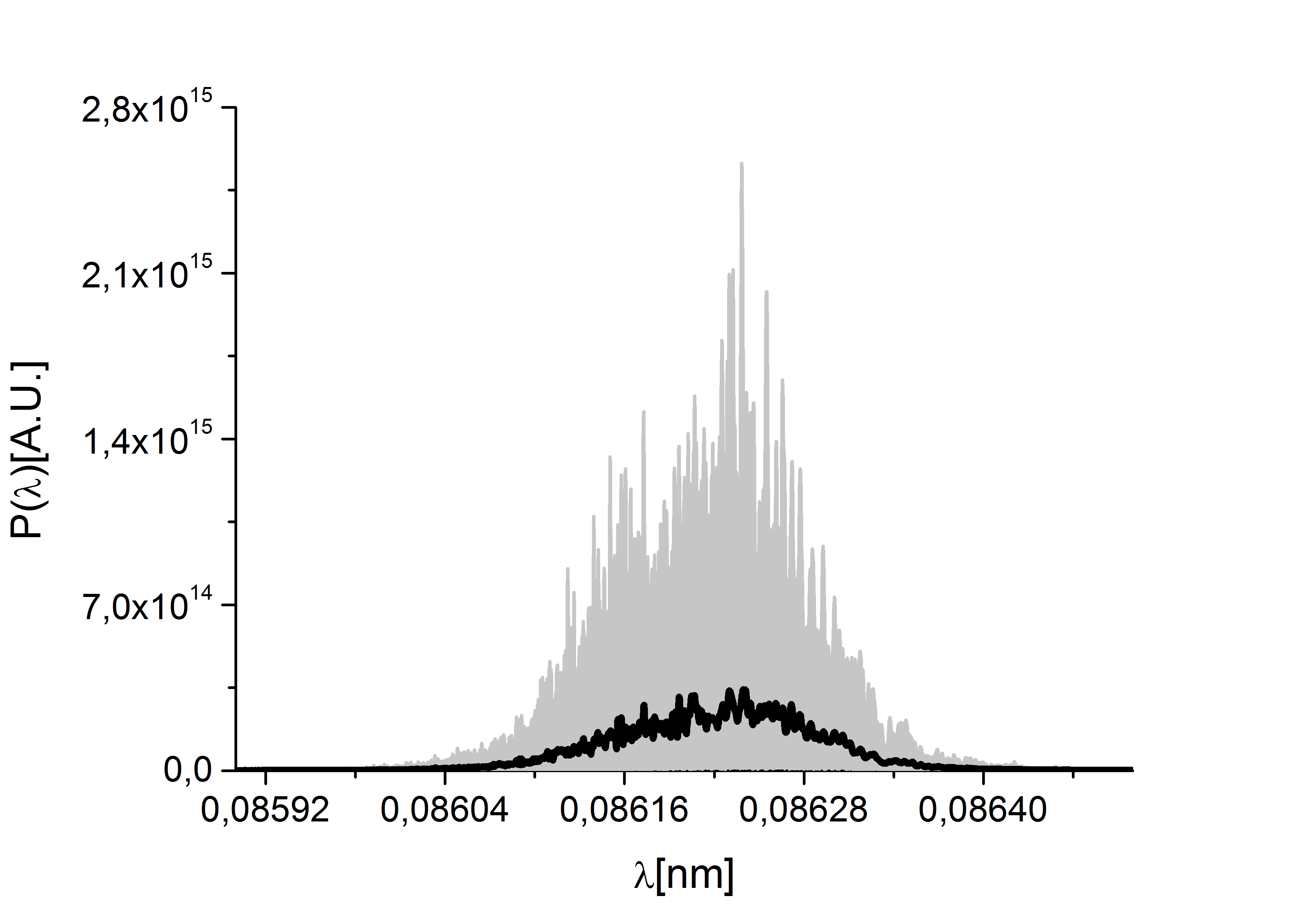}
\caption{Power and spectrum in the conventional SASE mode of operation at saturation, to be compared with Fig. \ref{pspouttap}. Grey lines refer to single shot realizations, the black
line refers to the average over a hundred realizations.} \label{sasesat}
\end{figure}
A comparison with the output power and spectrum for the conventional SASE mode at saturation can be made by inspecting Fig. \ref{sasesat}. The average peak power for the SASE pulse at saturation, Fig. \ref{sasesat}, is about $2 \cdot 10^{10}$ W with an average energy per pulse of about $0.25~$mJ. In the seeded case, Fig. \ref{pspouttap}, it reaches about $4 \cdot 10^{11}$ W with an average energy per pulse of about $3~$mJ.

This corresponds to an increase in flux from about $10^{11}$ photons per pulse to about $1.3 \cdot 10^{12}$ photons per pulse. Such increase, of about one order of magnitude, is due to tapering. Moreover, the final SASE spectrum relative bandwidth is $\Delta \lambda/\lambda \sim 1.4 \cdot 10^{-3}$ corresponding to about $20$ eV while, due to the enhancement of longitudinal coherence, the seeded spectrum has a FWHM relative bandwidth  $\Delta \lambda/\lambda \sim 7 \cdot 10^{-5}$, corresponding to about $1$ eV. This translates to an increase in spectral density of about a factor $20$.

Summing up, we obtain a bit more than one order of magnitude increase in peak power due to tapering, and more than an order of magnitude decrease in spectral width due to seeding. Combining the two effects, we obtain more than two orders of magnitude increase in spectral flux density from the SASE to the seeded-tapered case. Accounting for the high repetition rate of the European XFEL (27000 pulses per second), this correspond to about $10^{11}$ ph/s/meV for SASE case at saturation, compared to about $4 \cdot 10^{13}$ ph/s/meV in the case of the seeded-tapered case. Finally it should be noted that the background radiation flux at $7.2$ keV, mainly due to the first five segments of the radiator, see Fig. \ref{layoutund}, is limited to about two orders of magnitude less than the main output signal.

\begin{figure}
\includegraphics[width=0.50\textwidth]{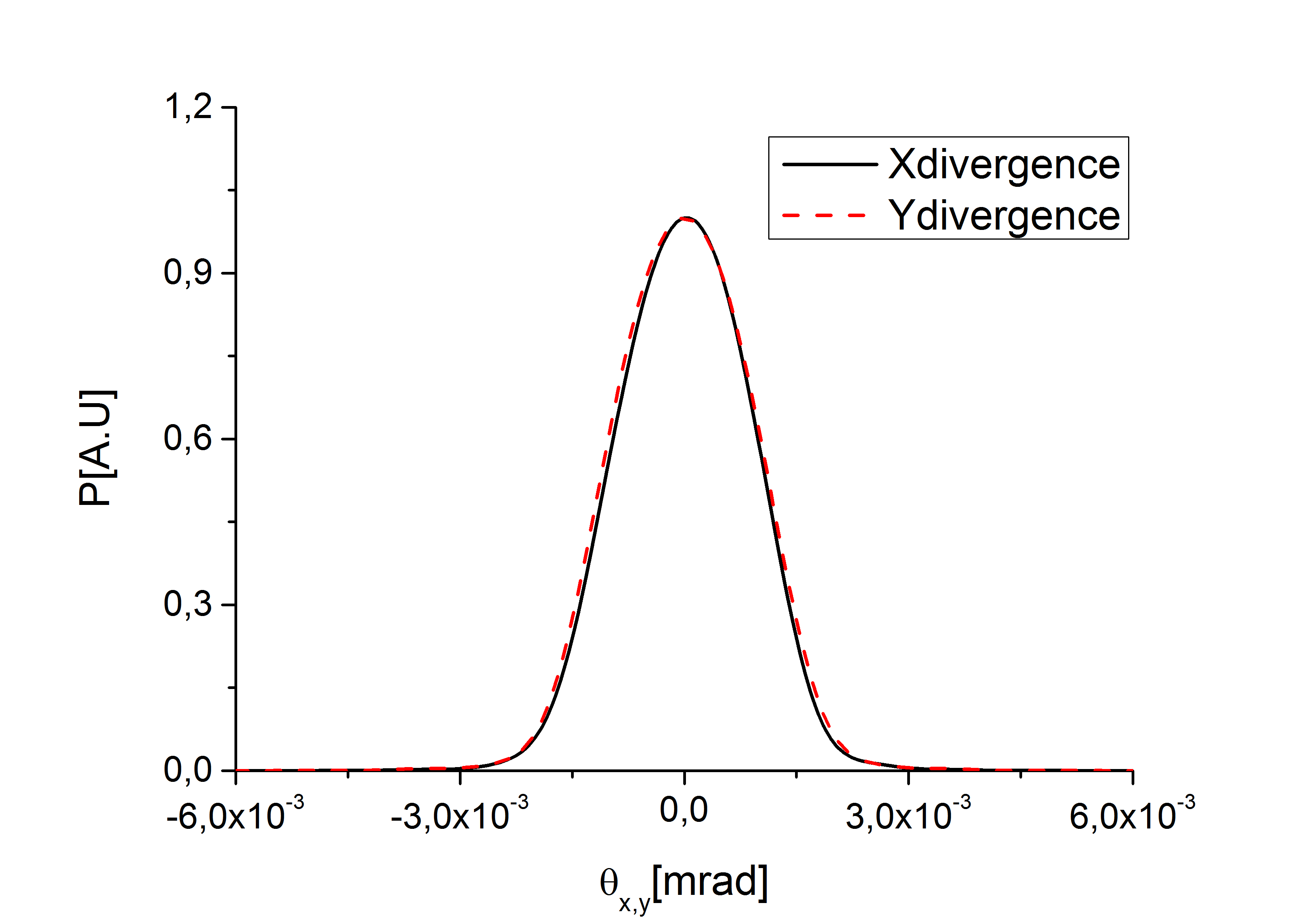}
\includegraphics[width=0.50\textwidth]{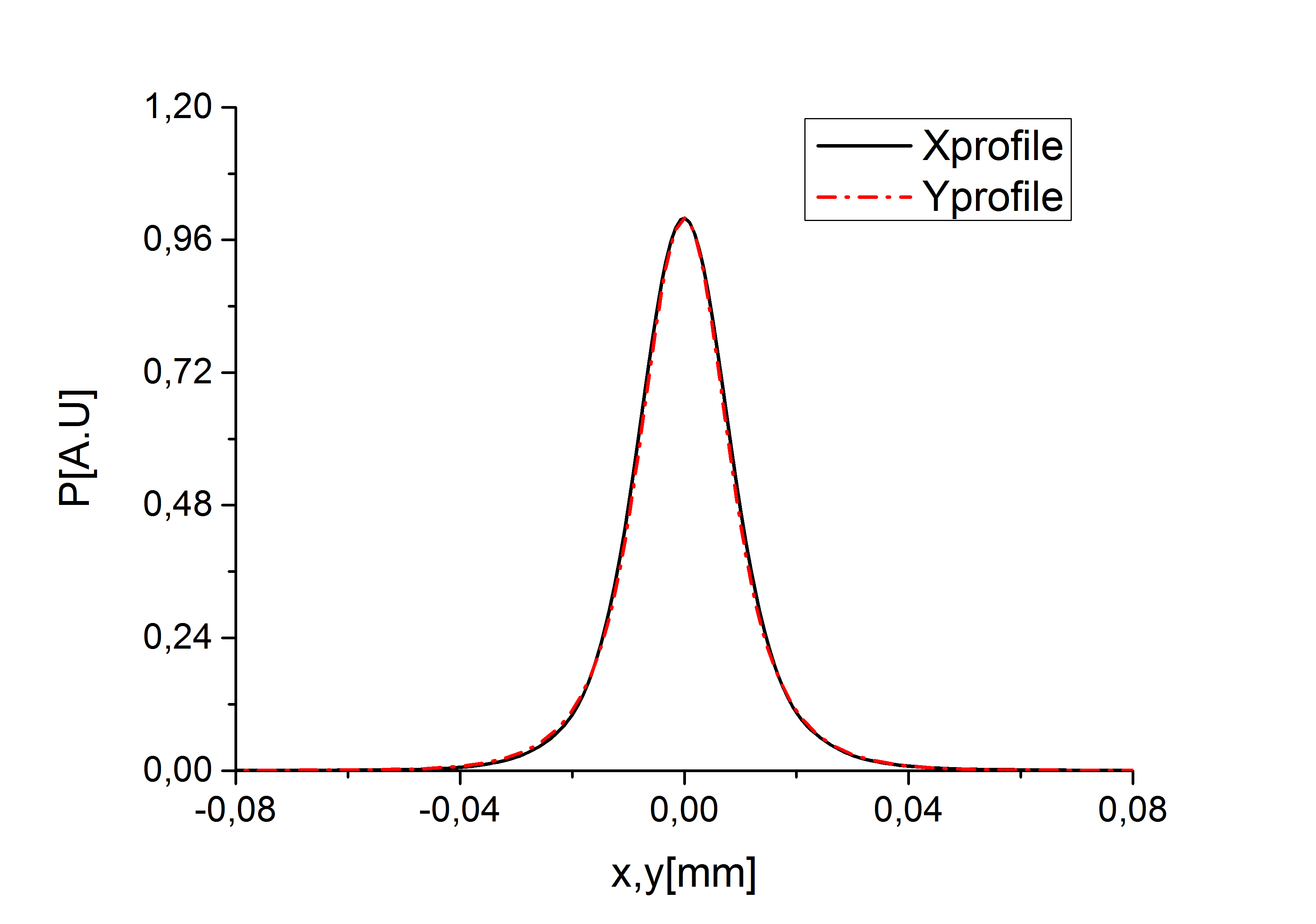}
\caption{Divergence and profile of the seeded FEL pulse at the exit of the output undulator.} \label{divprof}
\end{figure}
To close this section, in Fig. \ref{divprof}, we show size and divergence at the exit of the undulator, which are important in dealing with the beam transport through the X-ray beamline up to the experiment. The beam shape is about round with a FWHM size of about $20~\mu$m, and a FWHM divergence of about $2~\mu$rad.

\section{\label{sec:conc} Conclusions}

In this paper we investigated the potential of the European XFEL for experiments requiring high-average spectral flux and high photon energies like Nuclear Resonance Scattering (NRS) applications at $14.4$ keV. In particular, we  showed how a combination of Hard X-ray Self-Seeding (HXRSS), Coherent Harmonic Generation (CHG) and post-saturation tapering techniques, coupled with the high-repetition rate of the European XFEL can be advantageously used. HXRSS and tapering provide a combined increase in spectral flux of about two orders of magnitude, one to be ascribed to the bandwidth narrowing resulting from self-seeding, one due to tapering. CHG allows to overcome the decrease of HXRSS efficiency at high photon energy by exploiting the high-harmonic bunching in a HXRSS setup tuned at a sub-harmonic of the target energy, the second in our case,  without the introduction of any hardware change in the HXRSS setup. According to our start-to-end simulations this method should allow for a maximum flux of order $10^{13}$ photons per second in a meV bandwidth around $14.4$ keV, with a background at $7.2$ keV less than two orders of magnitude smaller. We underline that $10^{13}$ photons per second in a meV bandwidth at $14.4$ keV is more than two orders of magnitude larger than what would be achievable at the European XFEL in the nominal SASE mode at saturation, and three orders of magnitude larger than what is presently available at synchrotron radiation sources. We are thus confident that this method may constitute the key for  NRS experiments with ultra-high resolution of the order of $0.1$ meV.

\section{Acknowledgements}

We are grateful to  Massimo Altarelli, Serguei Molodtsov and Igor Zagorodnov for their interest
in this work.

\end{document}